\documentclass[floatfix,superscriptaddress,reprint,showpacs,aps,pra, nofootinbib]{revtex4-1}

\usepackage{graphics}
\usepackage{bm}
\usepackage{bbm}
\usepackage{amsmath}
\usepackage{amssymb}
\usepackage{graphicx}
\usepackage{amsthm}
\usepackage{comment}
\usepackage{mathtools}
\usepackage{bbold}
\usepackage[titletoc]{appendix}
\usepackage[dvipsnames]{xcolor}
\usepackage{courier}
\usepackage{hyperref}
\usepackage{float} 

\DeclareMathOperator{\PV}{PV}
\DeclareMathOperator{\Tr}{Tr}

\newcommand{\ket}[1]{|#1\rangle}

\newcommand{\ketbra}[2]{| \hspace{1pt} #1 \rangle \langle #2 \hspace{1pt} |}
\newcommand{\mbraket}[1]{\langle #1 \rangle}

\newcommand{\BC}{\mathcal{B}}

\newcommand{\HC}{\mathcal{H}}

\newcommand{\QC}{\mathcal{Q}}

\definecolor{url}{rgb}{0.25, 0.35, 0.7}
\definecolor{cite}{rgb}{0.33, 0.42, 0.18}
\definecolor{link}{rgb}{0.0, 0.48, 0.65}

\hypersetup{
    colorlinks=true,
    linkcolor=link,
    filecolor=magenta,
    citecolor=cite,
    urlcolor=url,
    pdftitle={Overleaf Example},
}

\begin{document}

\title{Quantum value for a family of \texorpdfstring{$I_{3322}$-like Bell}{} functionals}

\author{N.\ Gigena}
\email[]{nicolas.gigena@fuw.edu.pl}
\author{J.\ Kaniewski}
\email[]{jkaniewski@fuw.edu.pl}
\affiliation{Faculty of Physics, University of Warsaw, Pasteura 5, 02-093 Warsaw, Poland}

\begin{abstract}
We introduce a three-parameter family of Bell functionals that extends those studied in reference [Phys. Rev. Research {\bf 2}, 033420 (2020)] by including a marginal contribution. An analysis of their quantum value naturally splits the family into two branches, and for the first of them we show that this value is given by a simple function of the parameters defining the functionals. In this case we completely characterise the realisations attaining the optimal value and show that these functionals can be used to self-test any partially entangled state of two qubits. The optimal measurements, however, are not unique and form a one-parameter family of qubit measurements. The second branch, which includes the well-known $I_{3322}$ functional, is studied numerically. We identify the region in the parameter space where the quantum value can be attained, with two-dimensional systems and characterise the state and measurements attaining this value. Finally, we show that the set of realisations introduced in reference [Phys. Rev. A {\bf 82}, 022116 (2010)] to obtain the maximal violation of the $I_{3322}$ inequality succeeds in approaching the optimal value for a large subset of the functionals in this branch. In these cases we analyse and discuss the main features of the optimal realisations.
\end{abstract}

\maketitle

\section{Introduction}

A remarkable feature of quantum mechanics is the fact that quantum systems can exhibit correlations which are stronger than those predicted by any classical theory. J. Bell proved that this is the case in 1964 \cite{Be.64} by deriving an inequality that, if a local-realistic description of nature is assumed, imposes a constraint on the correlations between the outcomes of local measurements performed on distant systems. Bell's theorem provided a way to rule out the possibility of a local-realistic explanation for strong quantum correlations, a phenomenon nowadays referred to as {\it Bell nonlocality} that has become subject of increasing interest for the physics community, and in particular within the field of quantum information. Since Bell's seminal work several tools and techniques have been developed to further explore the nonlocal nature of quantum correlations, including the derivation of many new linear inequalities, generally known as {\it Bell inequalities} (see \cite{BC.14} for a review on Bell nonlocality).

Bell nonlocality is a quantum phenomenon which is not only relevant from the foundational point of view but it also has practical consequences: it plays a key role in device-independent quantum key distribution protocols and randomness certification \cite{BH.95, AG.97, GM.97, Co.11, AB.98, Pi.10, Ek.14, Pi.20, AM.16, BA.17}, and more generally device-independent quantum information processing \cite{BC.14, Br.14}. In a device-independent scenario the goal is to certify certain properties of quantum devices just by looking at the observed correlations, without making any assumptions about their physical description. Typically, in a device-independent scenario the conclusions are based on violating a certain Bell inequality, such as the CHSH inequality, for which it is known that observing the maximal quantum violation is sufficient to completely determine the state and measurements producing the observed correlations. For the case of the CHSH inequality the maximal violation can only be attained with anticommuting observables acting on a maximally entangled two-qubit state \cite{Ts.87, Ts.93, Su.87, PR.92}. This device-independent certification is known as {\it self-testing} \cite{Su.20, MY.04, MY.98} {\it or rigidity} \cite{RU.13}, and the fact that it can be obtained as a consequence of the maximal violation of a Bell inequality is not a coincidence. For probability points that can be realised with finite dimensional quantum systems it has been proved that only those being extremal points of the set of quantum correlations $\QC_{\rm finite}$ can be used to self-test both state and measurements \cite{GK.18}, and these extremal points can often be witnessed by maximally violating some Bell inequality over $\QC_{\rm finite}$. Beyond CHSH several other Bell inequalities proved to be useful for the purpose of self-testing \cite{SA.17, BA.20, BP.15, SA.16, Ka.16, AB.17}, amongst which those having a marginal contribution are the exception \cite{BP.15, WB.20, TF.21}. Recently a one-parameter family of Bell functionals was shown to exhibit a weaker form of self-testing \cite{Ka.20}, in which the maximal violation can be used to self-test the maximally entangled state of two qubits, but is not sufficient to fully determine the measurements.

In this work we propose to expand the family of functionals studied in Ref.~\cite{Ka.20} by including a marginal contribution which preserves their original symmetries, and analyse both their quantum value and the realisations giving rise to it. This analysis splits the new three-parameter family into two branches. In the first of these branches we derive an analytic expression for the quantum value as a function of the parameters defining the functionals and completely characterise the realisations attaining this value. These results show that the branch represents a natural extension of the parent family, in the sense that both the optimal value and realisations giving rise to it reduce to those found in Ref.~\cite{Ka.20} where the parameter introducing the marginal contribution vanishes. Moreover, we find that for these functionals weak self-testing statements can be made which are completely analogous to those of Ref.~\cite{Ka.20}. In this case, however, the quantum value self-tests a partially entangled pure two-qubit state which is a function of the parameters defining the Bell functional. Thus, this branch allows us to self-test any entangled two-qubit state. More specifically, for every two-qubit state we have a 1-parameter family of functionals.

The situation turns out to be very different in the second branch, for which a numerical analysis shows that in many cases the quantum value can only be reached with local Hilbert space of dimension $3$ or greater. There are nonetheless regions in the parameter space where the optimal value can be attained with two-qubit systems, for which we can give at least a realisation giving rise to it. In particular, for a large subset of these functionals we find that this value does not depend on one of the parameters. The remainder of the branch includes the $I_{3322}$ functional \cite{Fr.81, Sl.03, CG.04}, for which a particular sequence of finite dimensional realisations is known to numerically attain the quantum value in the limit of infinite local dimension~\cite{PV.10}. We study the performance of these special strategies beyond the $I_{3322}$ case and analyse the optimal solutions in those cases where they succeed in approaching the quantum value.

\section{A family of Bell functionals} \label{sec:2}

\subsection{Definition and symmetries}

We consider here a family of Bell functionals involving two parties, Alice and Bob, who perform local measurements that we label $x,y \in\{1, 2, 3\}$ respectively. A measurement $x$ performed by Alice yields an outcome $a\in\{1, -1\}$ and similarly a measurement $y$ by Bob yields an outcome $b\in\{1, -1\}$. We denote the probability to obtain outcomes $(a, b)$  when measurements $(x, y)$ are carried out by $p(ab|xy)$. We use $p(a|x)$ and $p(b|y)$ to denote the marginal distributions of Alice and Bob, respectively. Following the standard convention for Bell scenarios with binary outcomes we define the marginals and correlators as $\mbraket{A_x}=\sum_{a} a\, p(a|x)$, $\mbraket{B_y}=\sum_{b} b\, p(b|y)$ and $\mbraket{A_xB_y}=\sum_{ab} ab\, p(ab|xy)$ \cite{BC.14}. In terms of these quantities our functionals take the following form:
\begin{equation}
\begin{split} \label{eq:famber}
    \beta:=\,& \alpha_1[\mbraket{A_1}+\mbraket{A_2} +(-1)^{\alpha_2}\mbraket{B_1} +(-1)^{\alpha_2}\mbraket{B_2}]\\
    &+\mbraket{A_1 B_1} +\mbraket{A_1 B_2} +\mbraket{A_2 B_1} +\mbraket{A_2 B_2} \\
    &+ \alpha_3 \left[\mbraket{A_3 B_1} - \mbraket{A_3 B_2} + \mbraket{A_1 B_3} - \mbraket{A_2 B_3}\right],
\end{split}
\end{equation}
where $\alpha_2\in\{0, 1\}$ and $\alpha_1$, $\alpha_3\in\mathbb{R}$. Note that for $\alpha_2=0$ and $\alpha_1=\alpha_3=1$ the functional in Eq.~\eqref{eq:famber} becomes the well known $I_{3322}$ functional (see Appendix \ref{ap:B}), first introduced in Ref.~\cite{Fr.81} and further studied in Ref.~\cite{Sl.03} and \cite{CG.04}. 
It is not hard to check by looking at \eqref{eq:famber} that swapping parties, $A\leftrightarrow B$, leaves expression \eqref{eq:famber} invariant, and the same holds for the operation $A_1\leftrightarrow A_2$ ($B_1\leftrightarrow B_2$) followed by $B_3\rightarrow -B_3$ ($A_3\rightarrow -A_3$). As a direct consequence of these symmetries we can restrict our analysis to the case of $\alpha_1, \alpha_3$ being non negative. Indeed, a flip in the value of $\alpha_3$ can be cancelled by flipping the sign of $A_3$ and $B_3$, whereas a flip in  the sign of $\alpha_1$ can be cancelled by flipping the sign of all observables at once. For this reason, from now on, we will consider $\alpha_1, \alpha_3\geq 0$.

In the case $\alpha_1=0$ the expression in Eq.~\eqref{eq:famber} reduces to
\begin{eqnarray}\label{eq:corr}
\beta^{\rm cor}:=&&\;\mbraket{A_1 B_1} +\mbraket{A_1 B_2} +\mbraket{A_2 B_1} +\mbraket{A_2 B_2} \\
    &&+ \alpha_3 \left[\mbraket{A_3 B_1} - \mbraket{A_3 B_2} + \mbraket{A_1 B_3} - \mbraket{A_2 B_3}\right],\nonumber
\end{eqnarray}
a functional that has been exhaustively studied recently in Ref.~\cite{Ka.20} for $\alpha_3\in[0, 2]$, which turns out to be the interval where quantum realisations can give a value strictly larger than what is attainable with local-realistic strategies, i.e., where a quantum advantage can be obtained. The value attainable with local-realistic strategies is found to be $\beta^{\rm cor}_L=\max \{4, 4\alpha_3\}$, while the optimal value achievable with quantum realisations is $\beta^{\rm cor}_Q=4+\alpha^2_3$. Moreover, in the quantum case it is shown that the optimal value self-tests the maximally entangled state of two qubits and a one-parameter family of inequivalent two-qubit measurement arrangements. The functionals defined in Eq.~\eqref{eq:famber} can therefore be thought of as two inequivalent ways, depending on whether we set $\alpha_2$ to be $0$ or $1$, in which Eq.~\eqref{eq:corr} can be extended to a functional with a marginal contribution while preserving its symmetries. It is the purpose of this work to identify the regions in parameter space where the functional in Eq.~\eqref{eq:famber} provides a quantum advantage, and the quantum strategies optimising its value. We begin this study by computing the local and no-signalling values as functions of $\alpha_1$ and $\alpha_3$.

\subsection{Local and No-signalling values}

We will now find the maximal local $\beta_L$ and no-signalling $\beta_{NS}$ value for the functionals introduced in Eq.~\eqref{eq:famber}, as a function of parameters $\alpha_1, \alpha_3$, for both $\alpha_2=0$ and $\alpha_2=1$.

\begin{figure}[H] %8.8cm
\includegraphics[width=8.8cm,trim=0cm 0cm 0cm 0cm;clip]{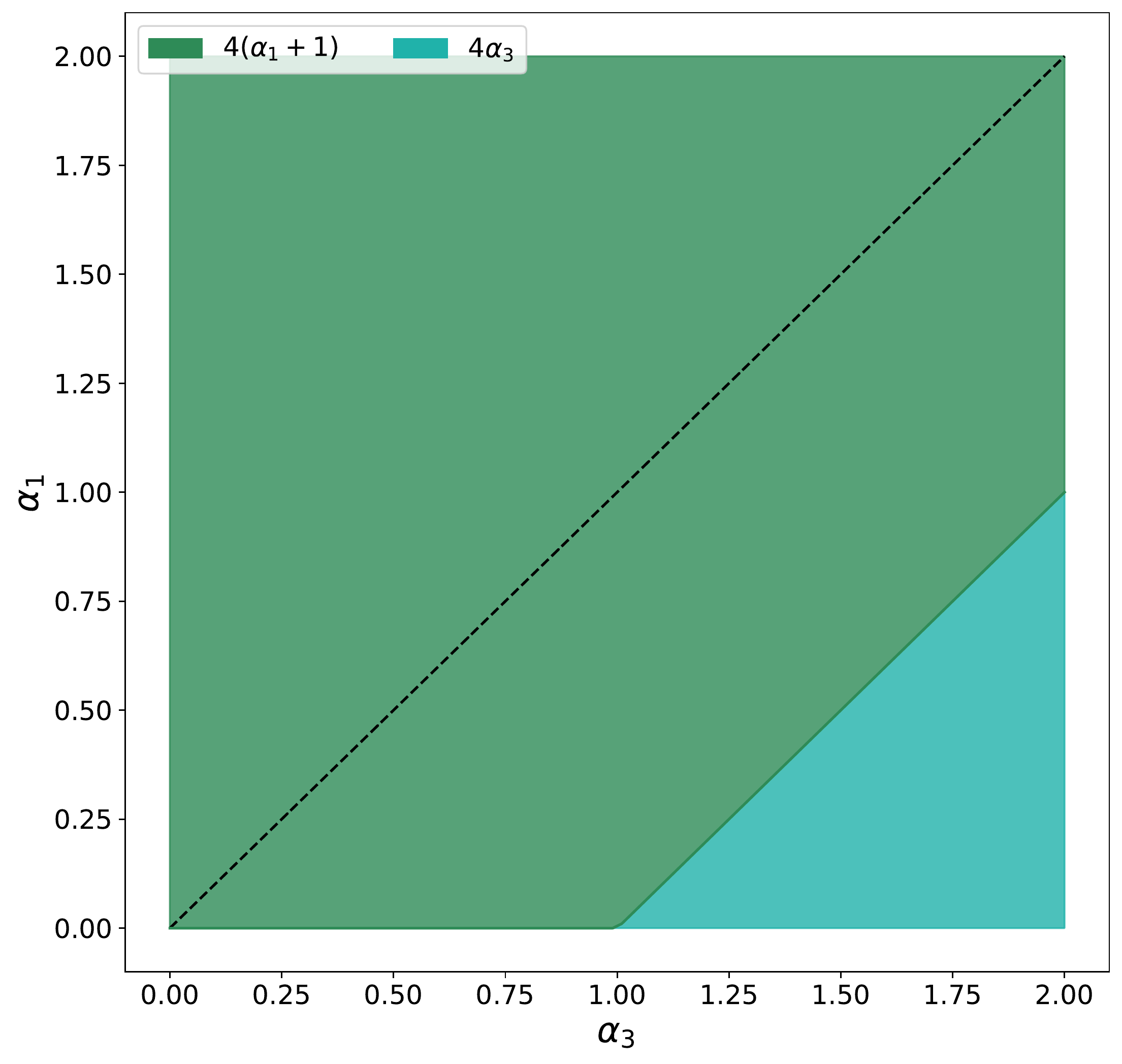}
\caption{Local value for the functionals in Eq.~\eqref{eq:famber} as a function of the parameters $\alpha_1$ and $\alpha_3$, for $\alpha_2=0$. The functionals in the region below the dashed line $\alpha_1=\alpha_3$ satisfy $\beta_{NS}>\beta_L$.}\label{FIG:1}
\end{figure}

\begin{figure}[h]
\includegraphics[width=8.7cm,trim=0cm 0cm 0cm 0cm;clip]{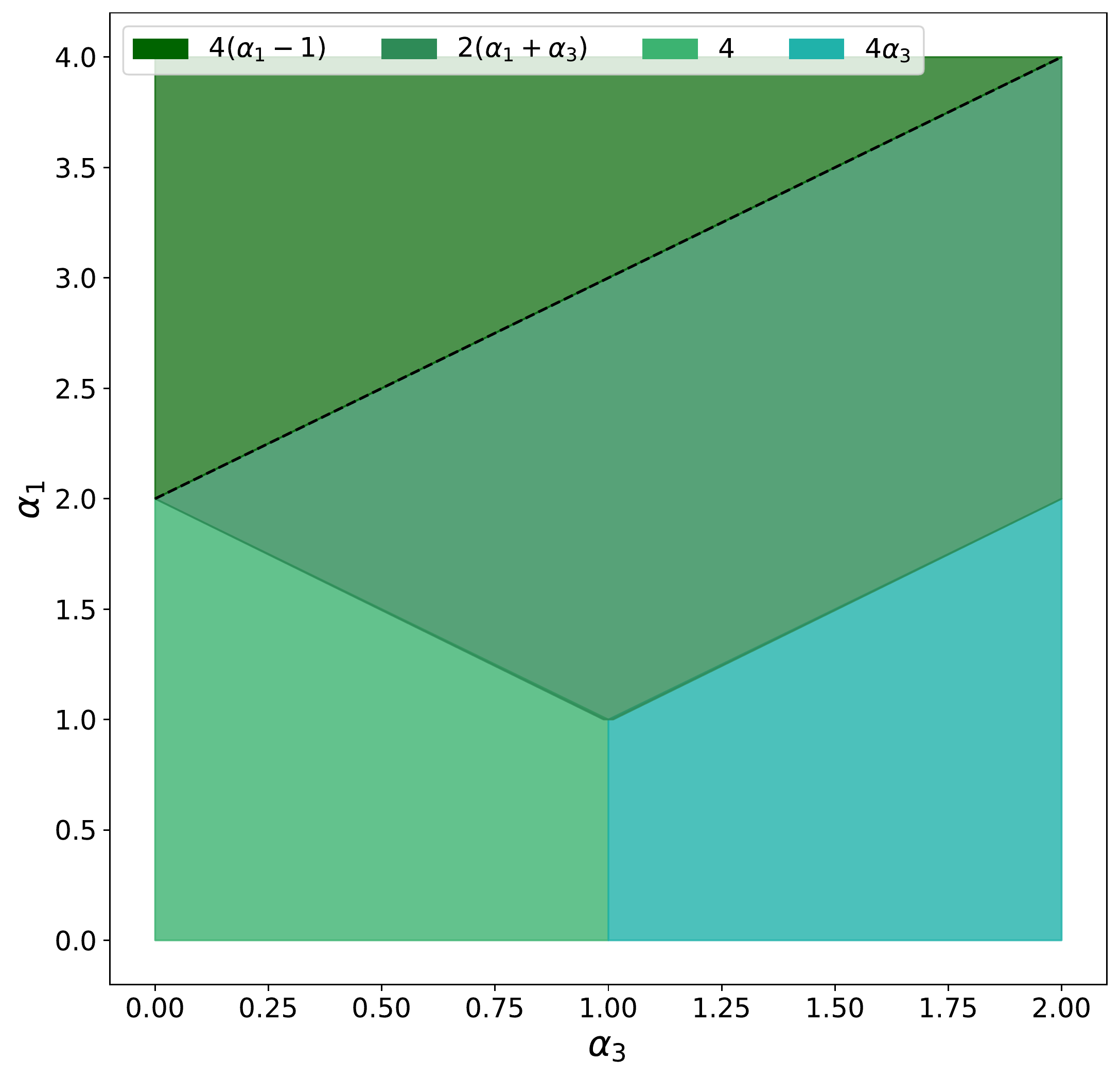}
\caption{Local value for the functionals in Eq.~\eqref{eq:famber} as a function of the parameters $\alpha_1$ and $\alpha_3$, for $\alpha_2=1$. The functionals in the region below the dashed line $\alpha_1=\alpha_3+2$ satisfy $\beta_{NS}>\beta_L$.}\label{FIG:2}
\end{figure}

Let us begin with $\alpha_2=0$. The symmetries described in the previous subsection make it easy to determine the local value of the functional, since we only need to care about the possible values of $\mbraket{A_1}\pm\mbraket{A_2}$ and $\mbraket{B_1}\pm\mbraket{B_2}$. For deterministic behaviours $\mbraket{A_1}+\mbraket{A_2}=\pm 2$ implies $\mbraket{A_1}-\mbraket{A_2}=0$ and $\mbraket{A_1}-\mbraket{A_2}=\pm 2$ implies $\mbraket{A_1}+\mbraket{A_2}=0$, with $B_{1(2)}$ satisfying analogous relations. Then, it is not hard to see that the local value in the case $\alpha_2=0$ is $\beta_L=4\alpha_3$ if $\alpha_1\leq\alpha_3-1$ or $\beta_L=4(\alpha_1+1)$ if $\alpha_1\geq\alpha_3-1$. Looking for the no-signalling (NS) value is also easy, since the extremal NS boxes in this scenario have been fully characterised \cite{BP.05, JM.05}. It turns out that the no signalling value is $\beta_{NS}=4(\alpha_1+1)$ if $\alpha_1\geq\alpha_3$ and $\beta_{NS}=4(\alpha_3+1)$ otherwise. The regions in the $\alpha_1, \alpha_3$ space in which the local value is given by the expressions derived above are shown in FIG. \ref{FIG:1}, where we also show the boundary for the region in which $\beta_{NS} \geq \beta_L$.
We can repeat this analysis for the case $\alpha_2=1$. By studying the possible values of $\mbraket{A_1}\pm\mbraket{A_2}$ and $\mbraket{B_1}\pm\mbraket{B_2}$ we conclude that the local value is $\beta_L=\max\{4(\alpha_1-1),\, 4\alpha_3,\, 2(\alpha_1+\alpha_3),\, 4\}$. Moreover, for the no signalling values we find $\beta_{NS}=\max\{\beta_L, 4(1+\alpha_3)\}$, which implies that $\beta_{NS}=\beta_L$ if $\alpha_1\geq \alpha_3+2$. The regions defined by these different local values, as well as the boundary of the region in which $\beta_{NS}>\beta_L$ are shown in FIG. \ref{FIG:2}.

\section{Quantum value for the \texorpdfstring{$\alpha_2=0$}{} family \label{sec:3}}

We are now interested in the maximal value attainable by the functionals in Eq.~\eqref{eq:famber}, for $\alpha_2=0$, with quantum systems. For a quantum realisation we can write the value of the functional as the expectation $\mbraket{W}=\Tr\; W\rho_{AB}$, over a quantum state $\rho_{AB}$, of the Bell operator
\begin{equation}
\begin{split}
    W=&\quad\alpha_1[(A_1+A_2)\otimes \mathbb{1}+ \mathbb{1}\otimes(B_1+B_2)] \\
    &+(A_1+A_2)\otimes(B_1+B_2) \\
    &+\alpha_3[(A_1-A_2)\otimes B_3+A_3\otimes(B_1-B_2)].
\end{split}
\end{equation}
Here the observables $A_x$ and $B_y$ are given by $A_x=M^A_{1|x}-M^A_{-1|x}$ and $B_y=M^B_{1|y}-M^B_{-1|y}$ respectively, with $M^A_{a|x}$ and $M^B_{b|y}$ measurement operators acting on the local Hilbert spaces $\HC_A$ and $\HC_B$ and determining outcome probabilities via the Born rule $p(ab|xy)={\rm Tr }\,M^A_{a|x}\otimes M^B_{b|y}\rho_{AB}$. For the region of interest a simple sum of squares (SOS) decomposition can be given for $W$. Indeed, if we assume $\alpha_1<\alpha_3$ and measurements to be projective, i.e., $A_x^2=B_y^2=\mathbb{1}$ $\forall\, x, y$,
it is straightforward to check that
\begin{equation}
    \nu\mathbb{1}\otimes\mathbb{1}-W=\sum_{i=1}^3 P_i^2 \label{eq:sos},
\end{equation}
where
\begin{eqnarray}
    \nu&=&2\left(\gamma^2+\frac{\alpha^2_3}{\gamma^2}\right)
    \label{eq:beta_Q}, \\
    \gamma&=&\sqrt{\frac{1}{2}(\alpha_3^2-\alpha_1^2)},
\end{eqnarray}
and the $P_i$'s are hermitian polynomials in the local observables given by the following expressions
\begin{eqnarray}
    P_1&=&\frac{1}{2\gamma}[  \alpha_1(A_1+A_2)\otimes \mathbb{1} + \alpha_3(A_1-A_2)\otimes B_3\notag  \\
    && \hspace{0.7cm}- 2\gamma^2\mathbb{1}\otimes\mathbb{1} ], \label{eq:P1}\\
    P_2&=&\frac{1}{2\gamma}[  \alpha_1 \mathbb{1}\otimes(B_1+B_2) + \alpha_3 A_3\otimes(B_1-B_2)\notag \\
    && \hspace{0.7cm}- 2\gamma^2 \mathbb{1}\otimes\mathbb{1} ], \label{eq:P2} \\
    P_3&=&\frac{1}{\sqrt{2}}[(A_1+A_2)\otimes\mathbb{1}-\mathbb{1}\otimes(B_1+B_2)] \label{eq:P3}.
\end{eqnarray}
By construction $\nu$ is an upper bound to the quantum value. For $\alpha_1 \leq \sqrt{\alpha_3^2+1}-1$ the resulting bound is actually tight and can be achieved by a quantum realisation $\{A_x, B_y, \ket{\Psi}\}$ where
\begin{equation} \label{eq:re1}
\begin{split}
\ket{\Psi}&=\cos\left(\frac{\varphi}{2}\right)\ket{00}+\sin\left(\frac{\varphi}{2}\right)\ket{11},\\
A_1&=B_1=\cos(\theta)\sigma_z+\sin(\theta)\sigma_x, \\
A_2&=B_2=\cos(\theta)\sigma_z-\sin(\theta)\sigma_x, \\
A_3&=B_3=\sigma_x.
\end{split}
\end{equation}
Here $\{\sigma_\mu,\, \mu=x, y, z\}$ are the Pauli matrices and the parameters $\theta$ and $\varphi$ satisfy the relations
\begin{eqnarray}
2\sin(\theta)&=&\sqrt{2\left(\gamma^2-\frac{\alpha_1^2}{\gamma^2} \right)},
\label{eq:sint} \\
\cos(\varphi)&=&\frac{\alpha_1}{\gamma^2}\cos(\theta), \label{eq:cosf} \\
\sin(\varphi)&=&\frac{\alpha_3}{\gamma^2}\sin(\theta). \label{eq:sinf}
\end{eqnarray}

\begin{figure}[h]
\includegraphics[width=8.7cm,trim=0cm 0cm 0cm 0cm;clip]{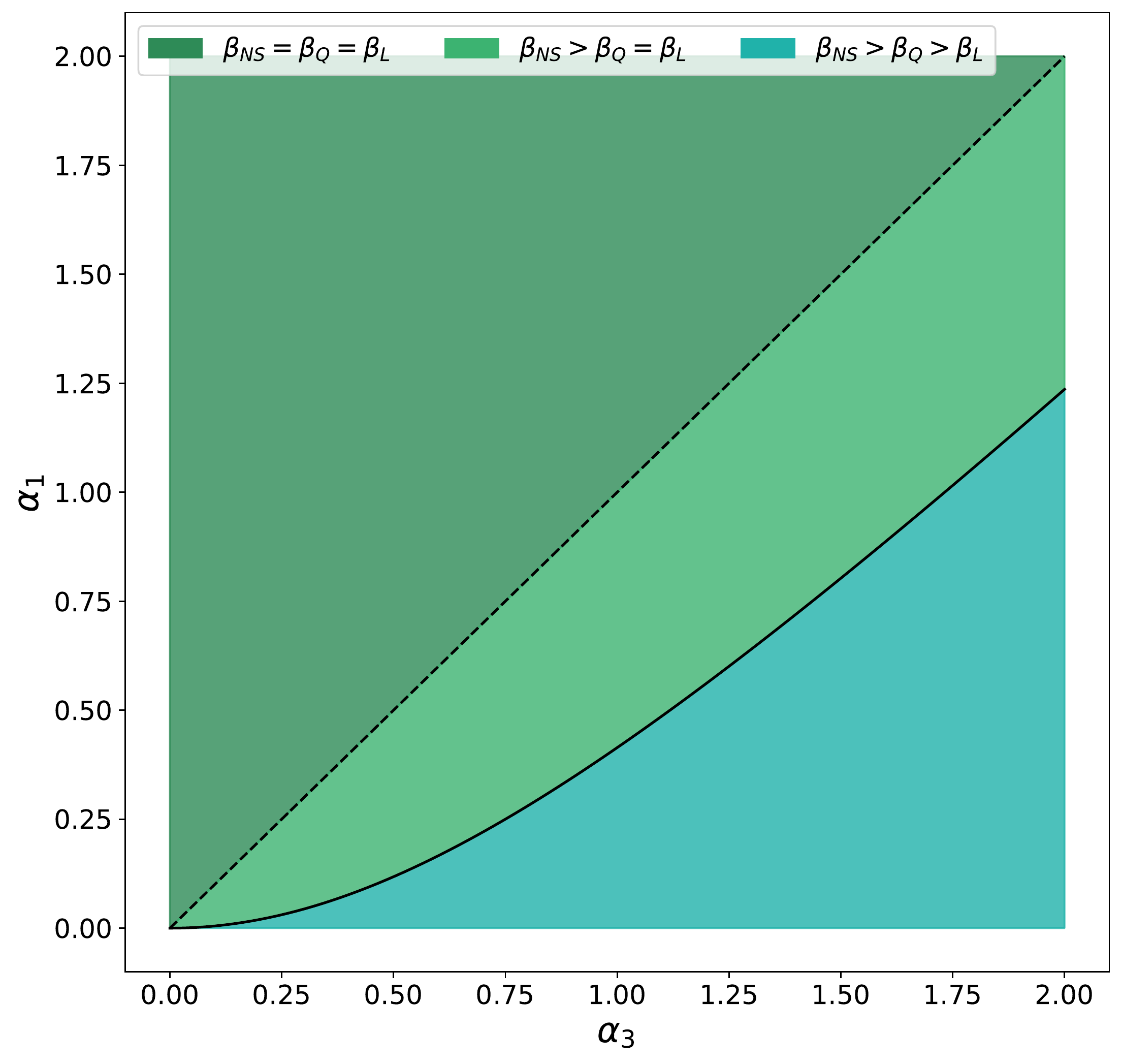}
\caption{Depiction of the quantum advantage ($\beta_Q>\beta_L$) region. The boundary of the region is given by the curve $\alpha_1=\sqrt{\alpha_3^2+1}-1$. As before, the region above the $\alpha_1=\alpha_3$ curve satisfies $\beta_{NS}=\beta_L$}\label{FIG:3}
\end{figure}

Eq.~\eqref{eq:sint} defines a region in parameter space where this solution is valid, and a direct calculation shows that this region is the one satisfying $\alpha_1\leq\sqrt{\alpha_3^2+1}-1$ (we consider here $\alpha_3\in[0, 2]$, see Appendix \ref{ap:A1} for a general result). Thus for functionals above the curve $f(\alpha_1,\alpha_3)=0$ with  $f(\alpha_1,\alpha_3)=1+\alpha_1-\sqrt{\alpha_3^2+1}$, the value $\nu$ cannot be attained with this realization. In fact, it is not hard to see that for these functionals $\nu$ is a strict upper bound. Indeed, let us write the Bell operator as
\begin{equation}
    W=\alpha_1 W_M+ W_C, \label{eq:wa1}
\end{equation}
with $W_M$ the marginal contribution to the functional and $W_C$ the correlation part, and write $\alpha_1=\epsilon+\tilde\alpha_1$, with $\tilde\alpha_1=\sqrt{\alpha_3^2+1}-1$. Plugging this expression in Eq.~\eqref{eq:wa1} and averaging we find
\begin{equation}
\begin{split}
    \mbraket{W}=\; &\tilde\alpha_1 \mbraket{W_M} + \mbraket{W_C}+\epsilon \mbraket{W_M} \\
    \leq \;& 4(\tilde\alpha_1+1) + \epsilon \mbraket{W_M}\leq 4(\alpha_1+1),
\end{split}
\end{equation}
where in the last line we used that on the curve $f(\alpha_1, \alpha_3)=0$ the quantum value is $4(\tilde\alpha_1+1)$. Thus we see that for $f(\alpha_1, \alpha_3)>0$ the local value $4(\alpha_1+1)$ becomes an upper bound for the quantum value, and since it can always be attained, it is the quantum value. The quantum value for the functionals is therefore $\beta_Q=\nu$ if $f(\alpha_1, \alpha_3)\leq 0$, and $\beta_Q=4(\alpha_1+1)$ otherwise. These values and the regions where they are attained are depicted in FIG. \ref{FIG:3}.

Combining these results implies that the region below the black solid curve in FIG. \ref{FIG:3} is the only one for which $\beta_Q>\beta_L$. For all these functionals we have been able to derive the upper bound $\nu$ in Eq.~\eqref{eq:beta_Q} and show that it is attainable with the realisation \eqref{eq:re1}, which implies this is indeed the optimal value $\beta_Q$. We are now interested in characterising the set of all quantum realisations achieving the optimal value in the aforementioned region.

Let $\{A_x, B_y, \rho_{AB}\}$ be a realisation achieving the value $\beta_Q$, and assume the marginal states $\rho_{A(B)}=\Tr_{B(A)}\,\rho_{AB}$ to be full-rank. We require the local states to have full-rank because otherwise we would not be able to characterise the local observables outside their support. Note first (see Appendix \ref{ap:A2}) that since the realisation is optimal the local observables must satisfy $A_x^2=B_y^2=\mathbb{1}$, i.e., the optimal measurements are projective. Moreover, note that from Eq.~\eqref{eq:sos} follows that saturating inequality $\Tr\,W\rho_{AB}\leq\beta_Q$ implies $\Tr\, P_i^2\rho_{AB}=0$ $\forall i$, and since both $\rho_{AB}$ and $P_i^2$ are positive semidefinite (recall that $P_i$ is hermitian) we see that $\rho_{AB}$ must have support on the intersection of the kernels of the $P_i$'s, i.e., $P_i\rho_{AB}=0$ $\forall i$. In Appendix \ref{ap:A2a} we show that this implies
\begin{eqnarray}
    \{A_1, A_2\}=\{B_1, B_2\}&=&2\left[ 1-\gamma^2+\frac{\alpha_1^2}{\gamma^2}  \right]\mathbb{1}, \label{eq:str1} \\
    \{A_1+A_2, A_3\}&=&\{B_1+B_2, B_3\}=0. \label{eq:str2}
\end{eqnarray}
Relations \eqref{eq:str1} and \eqref{eq:str2} turn out to be very useful to characterise the optimal realisation. Combined with Jordan's lemma \cite{Bha.97}, equation \eqref{eq:str1} on Alice's side implies that the local Hilbert space is of the form $\HC_A=\mathbb{C}^2\otimes\mathbb{C}^d$ for some $d\in\mathbb{N}$ and then, up to a unitary, the local observables $A_1$ and $A_2$ are
\begin{equation}\label{eq:A1A2}
\begin{split}
    A_1&=[\cos(\theta)\sigma_z+\sin(\theta)\sigma_x]\otimes\mathbb{1} \\
    A_2&=[\cos(\theta)\sigma_z-\sin(\theta)\sigma_x]\otimes\mathbb{1},
\end{split}
\end{equation}
where the angle $\theta$ is related to parameters $\alpha_1$ and $\alpha_3$ through equations \eqref{eq:sint}, \eqref{eq:cosf} and \eqref{eq:sinf}. In combination with Eq.~\eqref{eq:A1A2}, Eq.~\eqref{eq:str2} then implies that the most general form of $A_3$ is
\begin{equation}
    A_3=\sum_k[\cos(\mu_k)\sigma_x+\sin(\mu_k)\sigma_y]\otimes\ketbra{k}{k}, \label{eq:A_3}
\end{equation}
where $\mu_k\in[0, 2\pi)$ and $\{\ket{k}\}$ is an orthonormal basis of $\mathbb{C}^d$. The same derivation can be used to find Bob's observables with the same result. However, for reasons that will become clear soon it is convenient to write them in the following rotated form
\begin{equation}
\begin{split}
    B_1&=\sum_k[\cos(\theta)\sigma_z+\sin(\theta)(\cos(\nu_k)\sigma_x-\sin(\nu_k)\sigma_y)]\otimes\ketbra{k}{k}\\
    B_2&=\sum_k[\cos(\theta)\sigma_z-\sin(\theta)(\cos(\nu_k)\sigma_x-\sin(\nu_k)\sigma_y)]\otimes\ketbra{k}{k}\\
    B_3&=\sigma_x\otimes\mathbb{1},
\end{split}
\end{equation}
where as before $\nu_k\in[0, 2\pi)$ and $\{\ket{k}\}$ is an orthonormal basis of $\mathbb{C}^d$.

The block structure of the observables in the optimal realisation translates to the Bell operator, which we can write as
\begin{equation}
    W=\sum_{k, k'}R(\mu_k, \nu_{k'})\otimes\ketbra{k}{k}\otimes\ketbra{k'}{k'}, \label{eq:R}
\end{equation}
where $R(\mu_k, \nu_{k'})$ is the $2$-qubit Bell operator built up with the non-trivial part of the operators $\{A_x, B_y\}$ given above. The quantum value $\beta_Q$ is the largest eigenvalue of $W$, and equation \eqref{eq:R} shows that we can find that value in the spectra of the two-qubit operators $R(\mu_k, \nu_{k'})$.

It is not hard to prove (see Appendix \ref{ap:A2a}) that if $\beta_Q$ is an eigenvalue of $R(\mu, \nu)$ then $\mu=\nu$ must hold, and in that case the corresponding eigenspace is one-dimensional with the associated eigenstate being the one given in Eq.~\eqref{eq:re1}. Since this eigenstate is independent of the angle $\mu$ we conclude that any state $\rho_{AB}$ of a realisation attaining the quantum value $\beta_Q$ must be, up to a unitary, of the form
\begin{equation}
    \rho_{AB}=\Psi_{A'B'}\otimes\sigma_{\tilde A \tilde B},
\end{equation}
where $\Psi_{A'B'}$ is the density matrix of the state \eqref{eq:re1} and $\sigma_{\tilde A \tilde B}$ is an operator such that
\begin{equation}
    \Tr_{\tilde A \tilde B}\, W\,\mathbb{1}\otimes\sigma_{\tilde A \tilde B}=\sum_k p_k\, R(\mu_k,\mu_k),
\end{equation}
with $\sum_k p_k=1$.

We are now ready to completely characterise the probability points which attain the optimal value $\beta_Q$ for these functionals. A direct calculation gives:
\begin{equation}
\begin{split}
    \mbraket{A_{1(2)}}&=\mbraket{B_{1(2)}}=\cos(\theta)\cos(\varphi)\cos(\mu),\\
    \mbraket{A_1 B_1}&=\mbraket{A_2 B_2}=\cos^2(\theta)+\sin^2(\theta)\sin(\varphi)\cos(\mu),\\
    \mbraket{A_1 B_2}&=\mbraket{A_2 B_1}=\cos^2(\theta)-\sin^2(\theta)\sin(\varphi)\cos(\mu),\\
    \mbraket{A_3 B_1}&=-\mbraket{A_3 B_2}=\sin(\theta)\sin(\varphi), \\
    \mbraket{A_1 B_3}&=-\mbraket{A_2 B_3}=\sin(\theta)\sin(\varphi), \\
    \mbraket{A_3 B_3}&=\cos(\mu)\sin(\varphi). \label{eq:ppoint}
\end{split}
\end{equation}
We see from these expressions that the probability distributions saturating the functional value have a rather simple dependence on the angle $\mu$ parametrizing the whole family. In fact, it is apparent that they form a convex set with two extremal points, corresponding to $\mu=0$ and $\mu=\pi/2$. The set is therefore a line in probability space.

Combined with the relations in Eqs. \eqref{eq:sint}-\eqref{eq:sinf}, equations \eqref{eq:ppoint} also illustrate the trade-off between the advantage provided by the entanglement in the state and the contribution to the value of the marginal term in the functional. These expressions show that when $\alpha_1>0$ the marginal part of the functional can only contribute to the value if $\cos(\varphi)>0$, and the contribution will be larger as we let $\varphi$ approach $0$, which corresponds to a pure separable state in Eq.~\eqref{eq:re1}. We would then expect the optimal state to gradually become less entangled as we let the value of $\alpha_1$ increase. This is what we have found in our solution, since it is straightforward to verify that $\cos(\varphi)$ in Eq.~\eqref{eq:cosf} is an increasing function of $\alpha_1$.

It is worth mentioning that all the results derived above, including the self-testing statements, are easily seen to reduce to those reported in Ref.~\cite{Ka.20} in the case $\alpha_1=0$. In that sense, the $\alpha_2=0$ branch constitutes a natural extension of the parent family, with the advantage that the newly introduced functionals can be used to self-test {\it any} two-qubit pure entangled state.

\section{Quantum value for the \texorpdfstring{$\alpha_2=1$}{} family}

If we set $\alpha_2=1$ in functionals \eqref{eq:famber} and flip the sign of both $B_1$ and $B_2$ the associated Bell operator takes the form
\begin{equation}\label{eq:bopa1}
\begin{split}
    W=&\quad\alpha_1[(A_1+A_2)\otimes \mathbb{1}+ \mathbb{1}\otimes(B_1+B_2)] \\
    &-(A_1+A_2)\otimes(B_1+B_2) \\
    &+\alpha_3[(A_1-A_2)\otimes B_3+A_3\otimes(B_1-B_2)].
\end{split}
\end{equation}
As seen in this expression, the sign flip highlights the fact that these functionals, while inequivalent to those of the previous section, have the same symmetries. It can be seen already in the region plot in FIG. \ref{FIG:2} that choosing $\alpha_2=1$ in Eq.~\eqref{eq:famber} gives a very different set of functionals, for which we can expect the quantum value to have a more complicated dependence on parameters $\alpha_1$ and $\alpha_3$. It is worth noting that for the Bell operator \eqref{eq:bopa1} we can again write
\begin{equation}
    \nu\mathbb{1}\otimes\mathbb{1}-W=\sum_i P_i^2 \label{eq:sos2},
\end{equation}
where $\nu$ is given by again by \eqref{eq:beta_Q}, polynomials $P_{1(2)}$ are those of equations \eqref{eq:P1} and \eqref{eq:P2}, and $P_3$ reads
\begin{equation}
    P_3=\frac{1}{\sqrt{2}}[(A_1+A_2)\otimes\mathbb{1}+\mathbb{1}\otimes(B_1+B_2)].
\end{equation}
In this case, however, the numerical analysis described below suggests that the value $\nu$ is a strict upper bound for the whole region of interest.

\subsection{Numerical upper bounds and two-qubit values}

\begin{figure}[H]
\includegraphics[width=9cm,trim=0cm 0cm 0cm 0cm;clip]{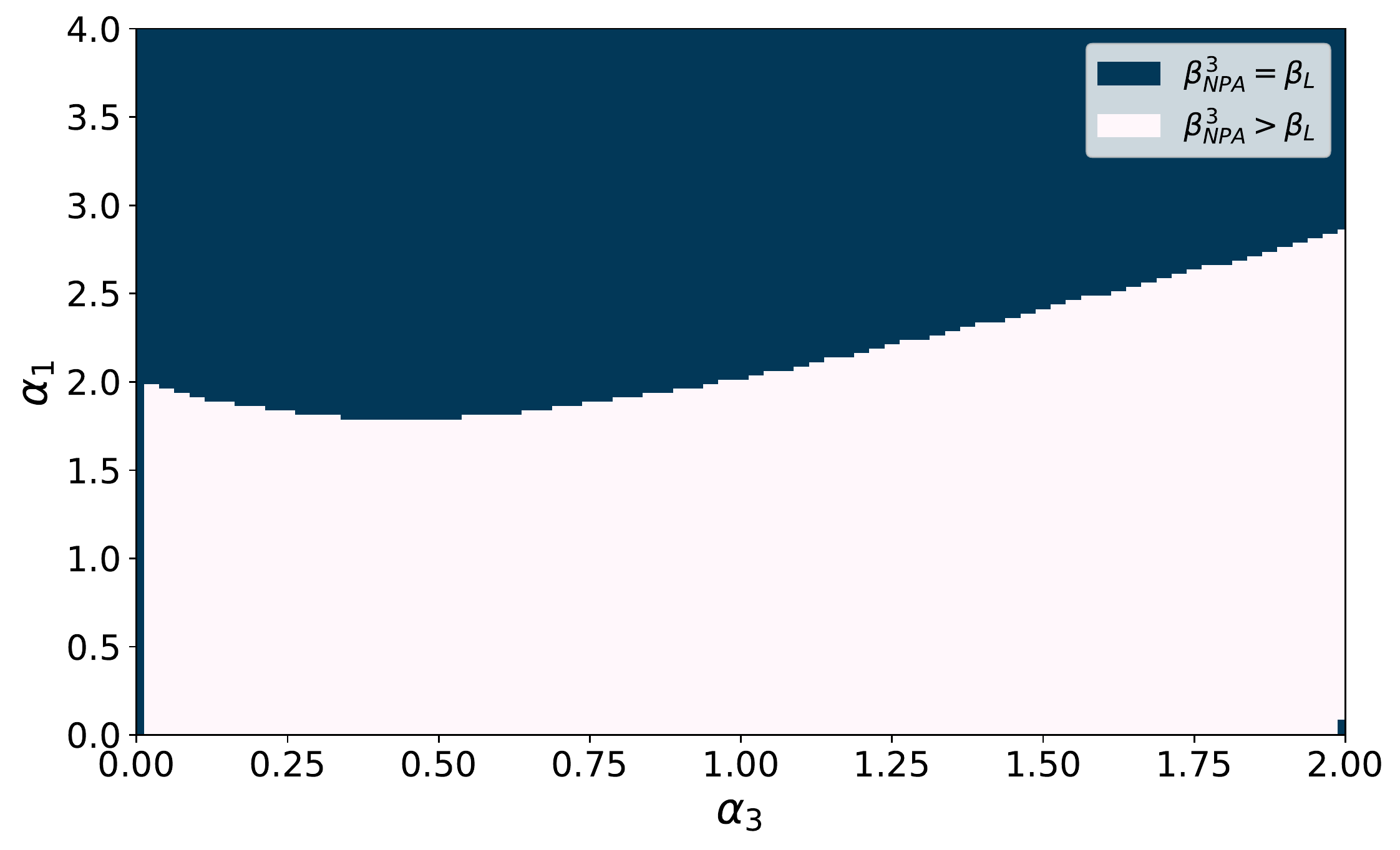}
\caption{Level 3 NPA upper bounds $\beta_{NPA}^{3}$ for the functionals \eqref{eq:famber} with $\alpha_2=1$. The quantum advantage ($\beta_{NPA}^3-\beta_L$) region is depicted in white.}\label{FIG:4}
\end{figure}

For given values of $\alpha_1$, $\alpha_3$ we can discard the possibility of saturating inequality $\mbraket{W}\leq \nu$ by finding tighter upper bounds. This can be done numerically using the NPA hierarchy of semidefinite programs \cite{NPA.07, NPA.08}, which we implemented using the convex-optimisation solver CVXPY \cite{DB.16}. We start by defining, within the region defined in FIG \ref{FIG:2}, a grid of $0.025\times 0.025$ such that each node corresponds to a particular functional in our family. Thus, for a total of 13041 functionals of the form \eqref{eq:famber} with $\alpha_2=1$, $\alpha_1\in[0, 4]$ and $\alpha_3\in[0, 2]$ we computed level 3 NPA upper bounds $\beta_{NPA}^{3}$ and identified the region where the upper bound is strictly greater than the local value $\beta_L$, which numerically we set as $\beta_{NPA}^{3}-\beta_L>10^{-6}$ . This region is depicted in FIG. \ref{FIG:4}.
It is seen in the figure that functionals satisfying $\alpha_1<\alpha_3$, for which the SOS decomposition is valid, fall within the region of quantum advantage. However a direct comparison shows that $\beta_{NPA}^{3}<\nu$ in the region below the $\alpha_1=\alpha_3$ line for $\alpha_1\neq 0$, implying that $\nu$ is a strict upper bound on the quantum value of all the functionals with a marginal term. For the functionals along the $\alpha_1=0$ line the solution is already known \cite{Ka.20} and it is a particular case of that found for the $\alpha_2=0$ case. Since this is a two-qubit solution, we can ask for which other functionals the quantum value can be attained with a two-qubit realisation.

\begin{figure}[h]
\includegraphics[width=9cm,trim=0cm 0cm 0cm 0cm;clip]{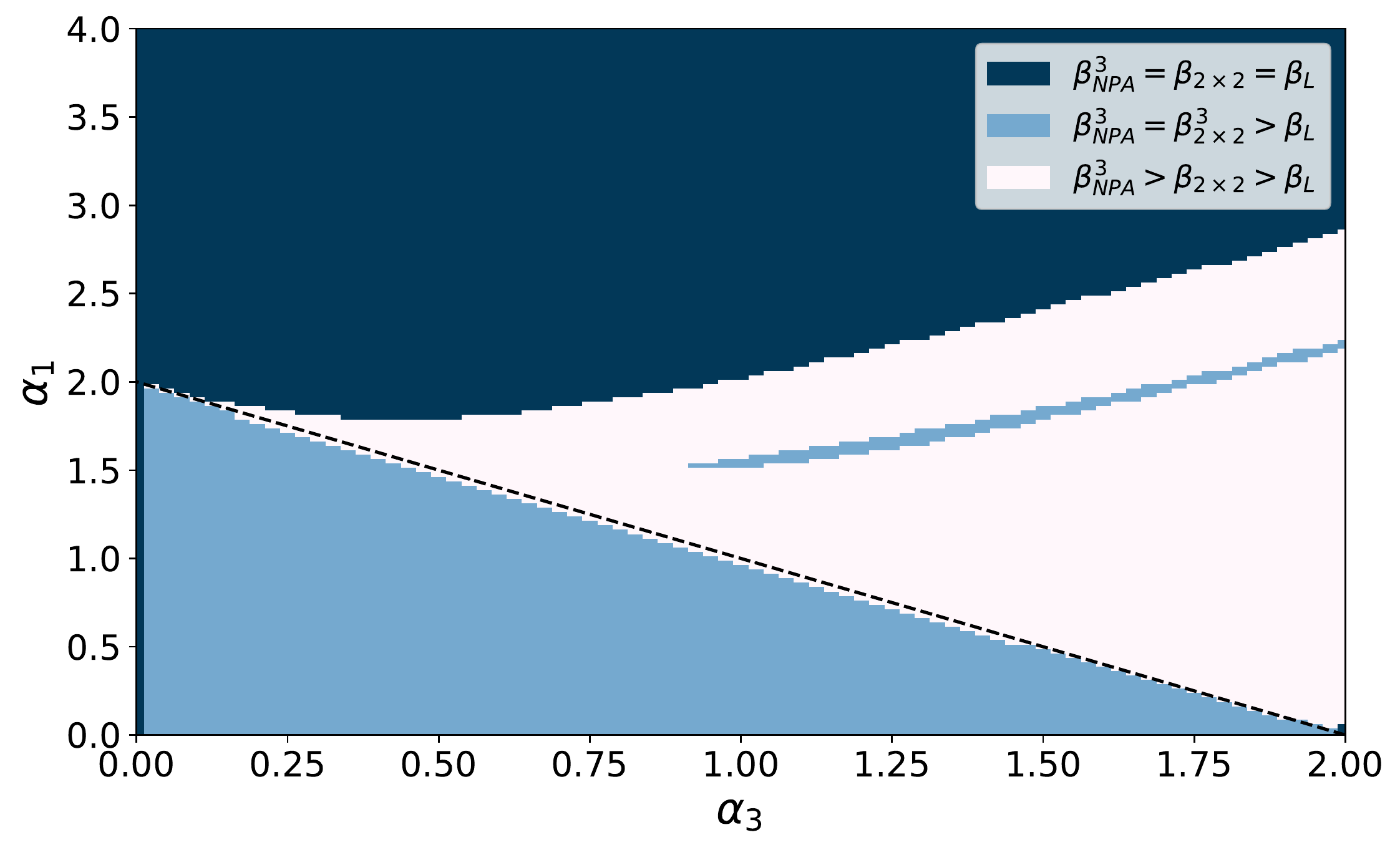}
\caption{Comparison of optimal two qubit values $\beta_{2\times 2}$ and NPA upper bounds $\beta_{NPA}^3$. The light blue regions show the functionals for which the quantum value is numerically attained with a two-qubit realisation. The dashed line corresponds to the $\alpha_1=-\alpha_3+2$ line, plotted only for reference.}\label{FIG:5}
\end{figure}

To compute the optimal two-qubit values we implemented a see-saw algorithm \cite{PV.10}, selecting the best value out of $150$ trials, $50$ iterations each. The results of this numerical optimisation are shown in FIG.~\ref{FIG:5}, in combination with the data relative to the local value $\beta_L$ and the NPA value $\beta_{NPA}^{3}$. The functionals in light-blue colour in the plot are those for which the see-saw procedure returns a value $\beta_{2\times 2}$ coinciding with the upper bound $\beta_{NPA}^{3}$ up to a $10^{-6}$ difference, while the white region shows those functionals for which $\beta_{NPA}^3-\beta_{2\times 2}>10^{-6}$. This condition is met, as shown in the figure, in two disjoint subregions.
For the almost triangular area below the $\alpha_1=-\alpha_3+2$ line we find that $\beta_{2\times 2}=4+\alpha_3^2$, which coincides with the optimal value for the $\alpha_1=0$ functionals. It is easy to check that this value can be attained with the two-qubit maximally entangled state $\ket{\Psi}=\frac{1}{\sqrt{2}}(\ket{00}+\ket{11})$ and operators
\begin{equation}
\begin{split}
A_1&=B_1=\cos(\theta)\sigma_z+\sin(\theta)\sigma_x, \\
A_2&=B_2=\cos(\theta)\sigma_z-\sin(\theta)\sigma_x, \\
A_3&=B_3=\sigma_x.
\end{split}
\end{equation}
This realization is clearly the $\alpha_1=0$ limit (with $\mu=0$) of that given in the previous section for the $\alpha_2=0$ case. In fact, not only does the two-qubit value match the upper bound $\beta_{NPA}^{3}$ but it already coincides with the $1+AB$ upper bound in the SDP hierarchy, which means that an SOS decomposition with polynomials in the $1+AB$ level exists for these functionals \cite{DW.08}. Unfortunately a numerical inspection of such decompositions shows them far from having a simple form and dependence on the parameters $\alpha_1$, $\alpha_3$, as was the case with the one we have encountered for the $\alpha_2=0$ subset. We speculate that the complexity of these decompositions comes from the fact that even though the functional depends on $\alpha_1$, this dependence is not reflected in the quantum value.

Above the $\alpha_1=-\alpha_3+2$ line, as depicted in FIG. \ref{FIG:5}, there is still a small subset of functionals for which the computed two-qubit values $\beta_{2\times 2}$ match the NPA upper bounds $\beta_{NPA}^{3}$. Our numerical analysis shows that all these functionals exhibit an interesting feature: the quantum value is attained by making one of the measurements trivial. Note that taking either $A_1 (B_1)$ or $A_2 (B_2)$ to be the identity operator implies that we can choose $\pm\mathbb{1}$ to be the optimal $B_3 (A_3)$, because this observable is completely determined by maximising the average $\mbraket{(A_1-A_2) \otimes B_3}$. As a result, our maximisation reduces to a scenario with two binary observables per party, for which it is known that the optimal value can always be attained with a two-qubit realisation \cite{AM.12}. Indeed, the results of our see-saw algorithm show that the optimal value is numerically attained with a partially entangled two-qubit state $\ket{\Psi}=\cos\left(\frac{\varphi}{2}\right)\ket{00}+\sin\left(\frac{\varphi}{2}\right)\ket{11}$ and operators
\begin{equation}
\begin{split}
A_1&=\:\mathbb{1}\qquad B_1=\cos(\theta)\sigma_z+\sin(\theta)\sigma_x,\\
A_2&=\sigma_z\qquad B_2=\cos(\theta)\sigma_z-\sin(\theta)\sigma_x,\\
A_3&=\sigma_x\qquad B_3=\mathbb{1}.
\end{split}
\end{equation}
We can build the Bell operator $W$ with these observables and compute its average over $\ket{\Psi}$, $\mbraket{W}_{\theta, \varphi}$. The quantum value is therefore
\begin{eqnarray}
    \beta_{2\times 2}=2\mathop{\max}_{\{\theta, \varphi\}}&\;&\mbraket{W}_{\theta, \varphi} \notag\\
    =2\mathop{\max}_{\{\varphi\}} &\,&\Bigg[ \frac{\alpha_1+\alpha_3}{2}+\frac{\alpha_1-\alpha_3}{2}\cos(\varphi) \\
    &+&\sqrt{[(\alpha_1-1)\cos(\varphi)-1]^2+\alpha_3^2\sin^2(\varphi)}\Bigg].\notag \label{eq:b22}
\end{eqnarray}
Note that for $\varphi\mapsto\pi$ we have $\ket{\Psi}\mapsto\ket{11}$, and in that case the optimal angle between $B_1$ and $B_2$ in the realization becomes $\theta=\pi$, thus recovering the local value $\beta_L$. 
As a final remark, it is worth mentioning that for these functionals the party swap symmetry $A\leftrightarrow B$ is clearly broken in the solution, implying the existence of inequivalent realisations attaining the optimal value.

The white subregion in FIG. \ref{FIG:5} shows the set of functionals for which the computed two-qubit values $\beta_{2\times 2}$ are strictly below the NPA upper bounds $\beta_{NPA}^{3}$ $(\beta_{NPA}^{3}-\beta_{2\times 2}>10^{-6})$. For studying higher dimensional realisations in this region a search based on the see-saw algorithm turns out to be very costly and time consuming. Instead, we focus on studying the performance of a particular set of realisations that we describe in the following subsection.

\subsection{P\'al-V\'ertesi realisations}\label{sec:PV}

As stated at the beginning of Sec. \ref{sec:2}, for $\alpha_2=0$ and $\alpha_1=\alpha_3=1$ Eq.~\eqref{eq:famber} becomes the well known $I_{3322}$ functional. This functional falls in the $\beta_{2\times 2}<\beta_{NPA}^{3}$ region, and in fact it has been shown that higher dimensional realisations can achieve values above the optimal two-qubit value $\beta_{2\times 2}^{I_{3322}}=5$ \cite{PV.10, DW.08, MT.13}\footnote{ Numerics have consistently shown that this is indeed the optimal two-qubit value. However, to the best of our knowledge an analytical proof for this statement is still lacking.}. The NPA upper bound we computed for this functional is $5.00350175$, figure that agrees up to $10^{-7}$ with that reported in Ref.~\cite{PV.10} where an extensive numerical analysis of the quantum value is presented. In this work Pál and Vértesi introduce a sequence of quantum realisations $\PV^{(n)}=\{A_x^{(n)}, B_y^{(n)}, \ket{\Psi^{(n)}}\}$, labelled by the the local Hilbert space dimension $n$, whose optimal value $\beta_{PV}^{(n)}$ approaches the NPA upper bound as $n$ increases. Based on the numerical evidence it is then conjectured that the quantum value is attained in the limit $n\rightarrow\infty$. In the following we show how to construct $\PV^{(n)}$ realisations (See Appendix \ref{ap:B} to see how these constructions relate to those presented in Ref.~\cite{PV.10}).

Let $n$ be odd and $\BC_A=\{\ket{k}, k=0, ..., n-1\}$ be a basis in which $A_1^{(n)}$ and $A_2^{(n)}$ are jointly block-diagonal as required by Jordan's lemma. Assume that $\ketbra{0}{0}$ is the only one-dimensional Jordan block in the decomposition, which implies that the complement of ${\rm span}\{\ket{0}\}$ is of the form $\mathbb{C}^2\otimes\mathbb{C}^d$, with $d=\frac{n-1}{2}$. Thus for $j=1, 3, ... n-2$ we can write $\{\ket{j}, \ket{j+1}\}=\{\ket{\uparrow, j}, \ket{\downarrow, j}\}$, so that operator $A_1$ takes the form
\begin{equation}
    A_{1}^{(n)}=\ketbra{0}{0}+\sum_{j=1}^{n-2}[\cos(\theta^A_j)\sigma_z + \sin(\theta^A_j)\sigma_x]\otimes\ketbra{j}{j}, \label{eq:A12}
\end{equation}
where parameters $\theta_j$ are the principal angles between $A_1$ and $A_2$. Operator $A_2$ takes the same form, but the off-diagonal elements have a negative sign. Now let $\ket{\bar k}\in\bar\BC_A$ denote the elements of $\BC_A$ sorted in decreasing order, i.e., $\ket{\bar k}=\ket{n-1-k}$. Then operator $A_3$ takes the form
\begin{equation}
    A_3^{(n)}=\ketbra{\bar 0}{\bar 0}+\sum_{j=1}^{n-2}\sigma_x\otimes\ketbra{\bar j}{\bar j}. \label{eq:A3}
\end{equation}
In an analogous way, if $\BC_B=\{\ket{k}, k=0, ..., n-1\}$ is the Jordan basis for $B_1$ and $B_2$, then operators $\{B_y\}$ are
\begin{eqnarray}
    B_{\substack{1}}^{(n)}&=&\ketbra{0}{0}+\sum_{j=1}^{n-2}[\cos(\theta^B_j)\sigma_z+\sin(\theta^B_j)\sigma_x]\otimes\ketbra{j}{j}\qquad \label{eq:B12} \\
    B_3^{(n)}&=&\ketbra{\bar 0}{\bar 0}+\sum_{j=1}^{n-2}\sigma_x\otimes\ketbra{\bar j}{\bar j} \label{eq:B3}, 
\end{eqnarray}
with $B_2$, as in the case of $A_2$, being the same as $B_1$ with negative off-diagonal coefficients. Finally, the state $\ket{\Psi^{(n)}}$ in the $\PV^{(n)}$ realisation is given by
\begin{equation}
    \ket{\Psi^{(n)}}=\sqrt{\lambda_{0}}\ket{ 0}\ket{\bar 0}+\sum_{\substack{j=1\\a=\uparrow, \downarrow}}^{n-2}\sqrt{\lambda_{a j}}\ket{a, j}\ket{\bar a, \bar j}. \label{eq:pvst}
\end{equation}

For even local dimension the observables have the same block-diagonal structure, but the symmetry is lost. $B_1$, $B_2$ and $A_3$ have $\frac{n}{2}$ $2$-dimensional blocks, whereas for $A_1$, $A_2$ and $B_3$ the first and last blocks have dimension $1$.

\begin{figure}[h]
\includegraphics[width=8.75cm,trim=0cm 0.5cm 0cm 0cm;clip]{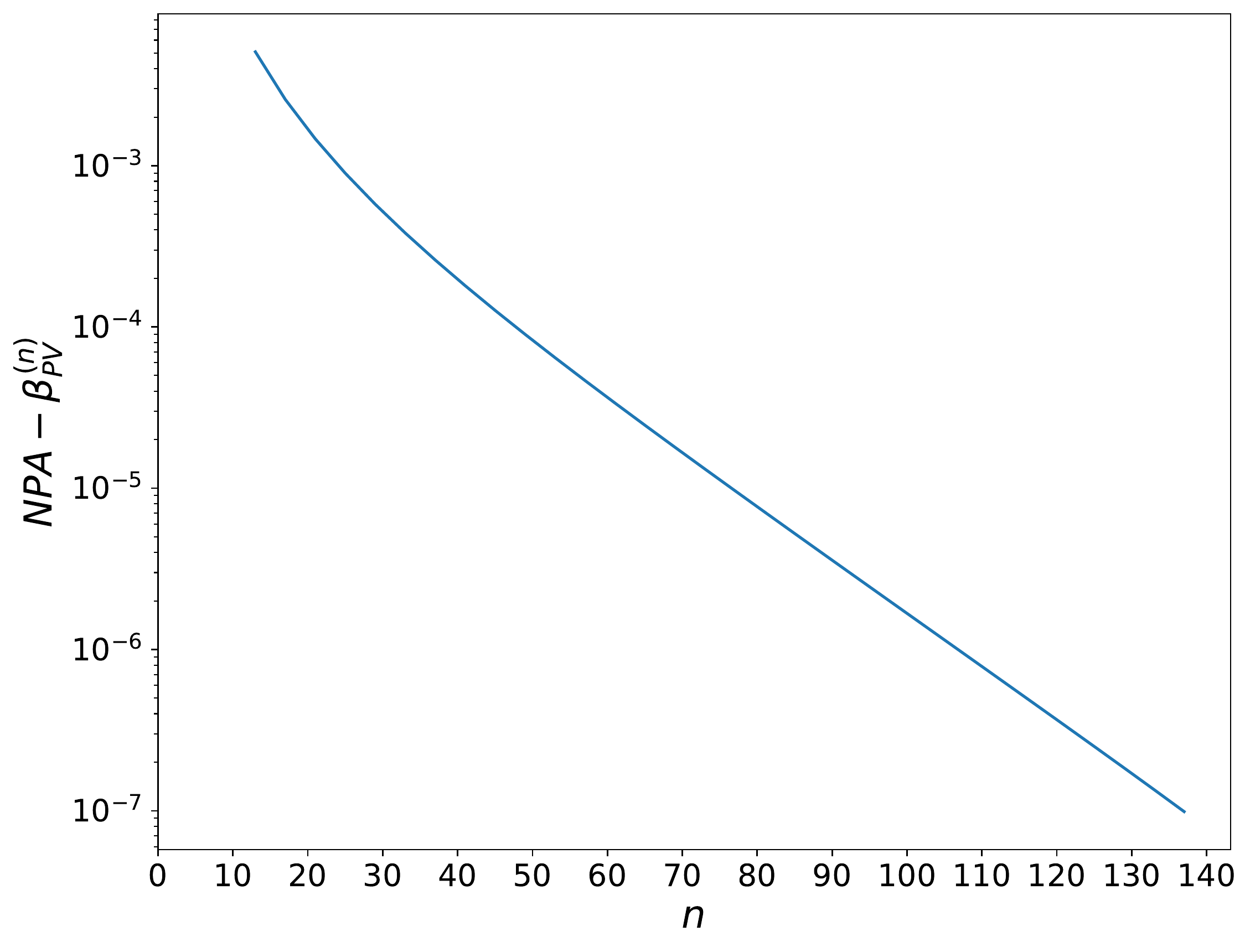}
\caption{Distance of the optimal value $\beta_{PV}^{(n)}$, achieved with an $n$-dimensional $\PV^{(n)}$ realisation, to the level $3$ NPA value as a function of local dimension $n$.}\label{FIG:log}
\end{figure}

\begin{figure}[b]
\includegraphics[width=9cm,trim=0.75cm 0.5cm -0.6cm 0cm;clip]{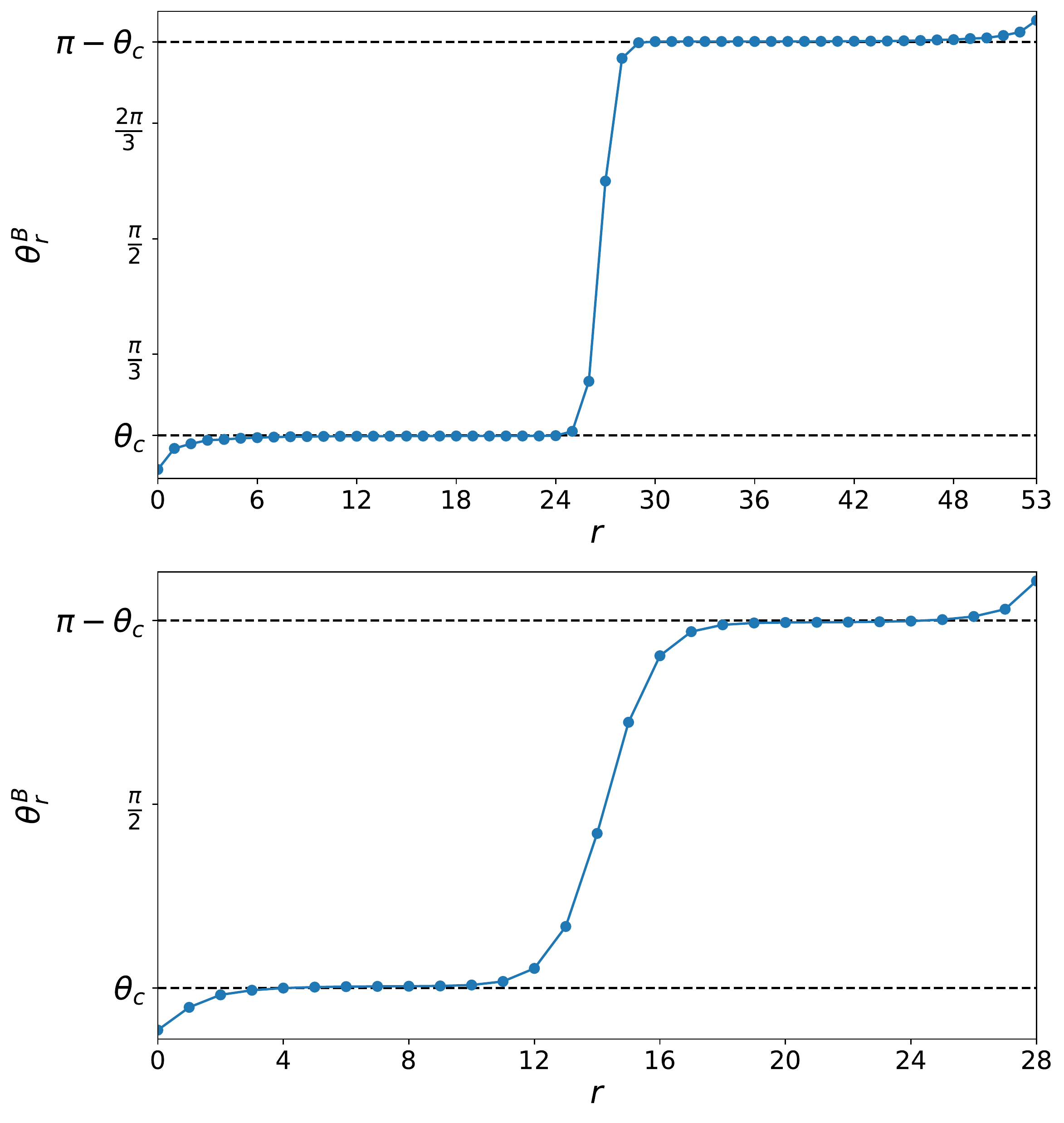}
\caption{Angles for Bob's observables in the optimal PV realisation for the $(\alpha_1=0.75, \alpha_3=1.3)$ functional (top) and for the $(\alpha_1=0.75, \alpha_3=1.85)$ functional (bottom).}\label{FIG:9}
\end{figure}

As seen in these expressions, the measurement operators on both sides are completely parametrized by the principal angles $\{\theta_j^A\}$ and $\{\theta_j^B\}$. The optimal value of the average $\mbraket{W}$ attainable with a $\PV^{(n)}$ realisation is therefore a function of these angles and the Schmidt coefficients of the state \eqref{eq:pvst}. We used the \texttt{scipy.optimize} package to compute the maximal value of this function for different local Hilbert space dimensions, using as initial guess a set of values that resembles the solution presented in Ref.~\cite{PV.10} (see Fig.~\ref{FIG:9}). Our results, depicted in FIG.~\ref{FIG:log}, reproduce those reported by Pál and Vértesi with an optimal value converging to $\beta_{PV}=5.00350154$, which is below the NPA upper bound by less than $10^{-6}$.
It is worth noting that the principal angles in the optimal solution are the same for both parties, which should not be surprising as the functional is symmetric with respect to swapping Alice and Bob. We used this observation in the optimisation function in order to only optimise over Alice's angles, which reduces considerably the running time.

\begin{figure}[h]
\includegraphics[width=9cm,trim=0cm 0.25cm 0cm 0.25cm;clip]{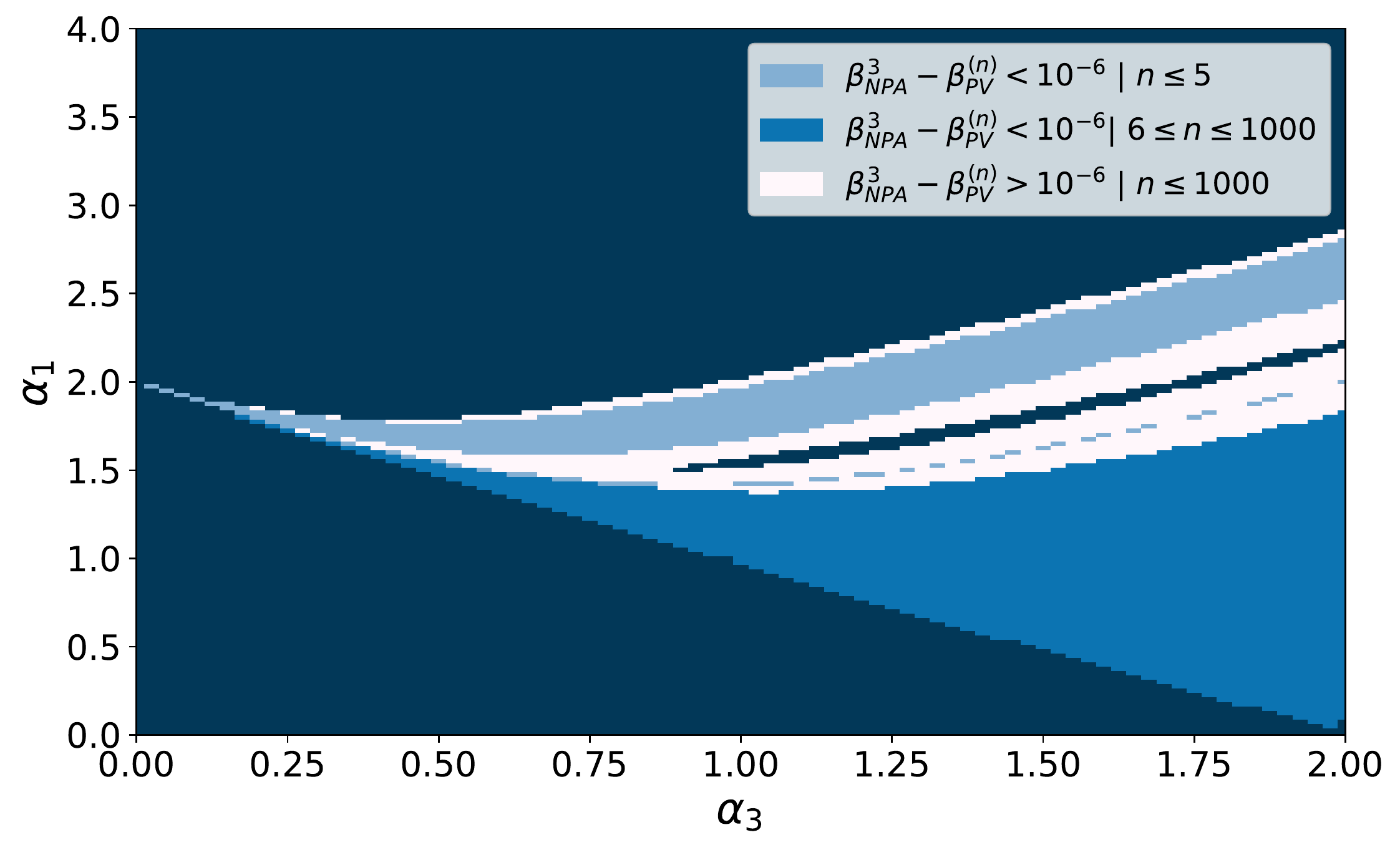}
\caption{Optimal values $\beta_{PV}^{(n)}$ attained with $\PV^{(n)}$ realisations for the functionals in the region of interest, compared with their respective level $3$ NPA upper bounds $\beta_{NPA}^{3}$. For the functionals plotted in light blue the gap between two-qubit values and the NPA upper bound is closed already at dimension $3$ or $5$.}\label{FIG:7}
\end{figure}

After reproducing the results previously reported for $I_{3322}$ we moved onto exploring the values that other functionals in our family can attain when the observables are constrained to have this particular form. Thus, for each functional in the white region of FIG.~\ref{FIG:5} we computed a sequence of optimal values $\beta_{PV}^{(n)}$ attained with a $\PV^{(n)}$ realisation, in the same way we did for the $I_{3322}$ case. The sequence starts at local dimension $n=3$ which increases with step $1$ up to $n=50$. The step then changes to $10$ until $n=150$, where it increases again to $30$ up to $n=300$ and then to $50$ until $n=600$. Above $n=600$ the step is $100$. This particular selection of steps is aimed to reduce the computation time and results form an analysis of the values observed in a sample of functionals within the region. Each value $\beta_{PV}^{(n)}$ computed in this way for a given functional is compared with both the previous one (if $n>3$) and the upper bound $\beta_{NPA}^3$. The sequence is then stopped either if $\beta_{NPA}^3-\beta_{PV}^{(n)}<10^{-6}$ or if $5$ consecutive values are the same up to $10^{-7}$: In the first case we consider the gap to be closed, while in the second we consider that the sequence converged to its maximal value even if it does not agree with the upper bound. The results of this numerical search are depicted in FIG. \ref{FIG:7}.
As can be seen in the figure, for most of the functionals a $\PV^{(n)}$ realization closes the gap between the two-qubit value and the NPA upper bound. As before, the white region in the plot shows the functionals for which the distance between the attained value and the upper bound is greater than $10^{-6}$. Amongst the points where $\beta_{NPA}^{3} - \beta_{PV}^{(n)}<10^{-6}$ for some $n$, the ones with the largest value for $\alpha_1$ reach the optimal value already at dimension $3$ or $5$. These are shown in light blue on the plot. It is worth noting that convergence to the optimal value for these functionals is quite fast as compared with the results we obtained from the see-saw algorithm, even after increasing the number of random starting points to $200$. For the remainder of the set it is seen that the minimal local dimension at which the gap is closed increases up to $n=1200$ as the value of $\alpha_1$ decreases.
\begin{figure}[b]
\includegraphics[width=9cm,trim=0cm 0.25cm 0cm 0cm]{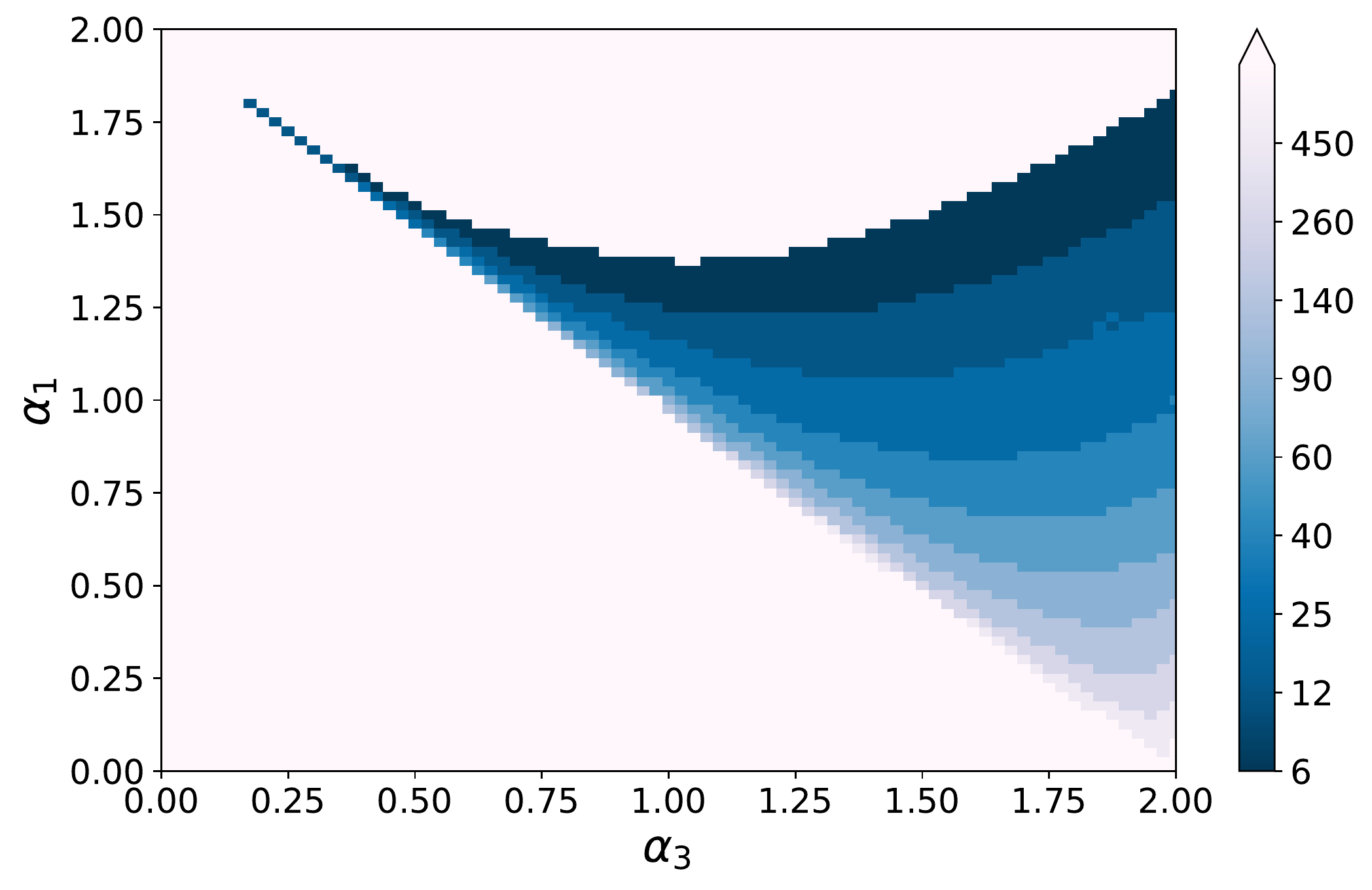}
\caption{Functionals for which the quantum value can be approached with $\PV^{(n)}$ realisations with local dimension $n>5$. The colour bands show the interval containing the lowest dimension for which $\beta_{NPA}^{3}-\beta_{PV}^{(n)}<10^{-6}$. For clarity, we only include in the plot the functionals labeled as $\beta^3_{NPA}-\beta_{PV}^{(n)}<10^{-6} | \,6\leq n\leq 1000$ in FIG. \ref{FIG:7}.}\label{FIG:8}
\end{figure}
In FIG. \ref{FIG:8} we illustrate this dependence by partitioning the interval $[6, 1200]$ in 9 different bands with the first one starting at dimension $6$ and the last one starting at dimension $450$.

In FIG.~\ref{FIG:9} we show, as an example, the angles parametrizing the optimal measurement for the functionals defined by $(\alpha_1=0.75, \alpha_3=1.3)$ and $(\alpha_1=0.75, \alpha_3=1.85)$. Note first that for $\alpha_3=1.85$ the number of principal angles $m$ in the solution is lower than in the $\alpha_3=1.3$ case as is expected from FIG.~\ref{FIG:8}.
The curves have nonetheless a similar behaviour, which is observed in all these solutions: it starts at a value $\theta_0$ below $\frac{\pi}{2}$ and quickly rises to a slightly greater value $\theta_c$ which remains almost constant until, in the vicinity of $\frac{m}{2}$, it rises to $\frac{\pi}{2} - \theta_c$ in a short interval and finally to $\frac{\pi}{2}-\theta_0$. For a fixed value of $\alpha_1$ we consistently observe that the almost constant value $\theta_c$ increases as $\alpha_3$ increases. Moreover, as $\alpha_3$ increases we also observe the zone of transition from $\theta_c$ to $\frac{\pi}{2}-\theta_c$ to become wider, involving a larger subset of the principal angles. Finally, it is observed that the behaviour in the extremes of the curve is essentially the same in all cases, regardless of the dimension of the solution. Such behaviour might be an artifact of computing the optimal value of a finite dimensional realisation and we do not expect it to be present in the infinite dimensional limit in which $\beta_{PV}^{(n)}$ is expected to converge to the quantum value.

\begin{figure}[h]
\includegraphics[width=9cm,trim=0cm 0.5cm 0cm -0.25cm;clip]{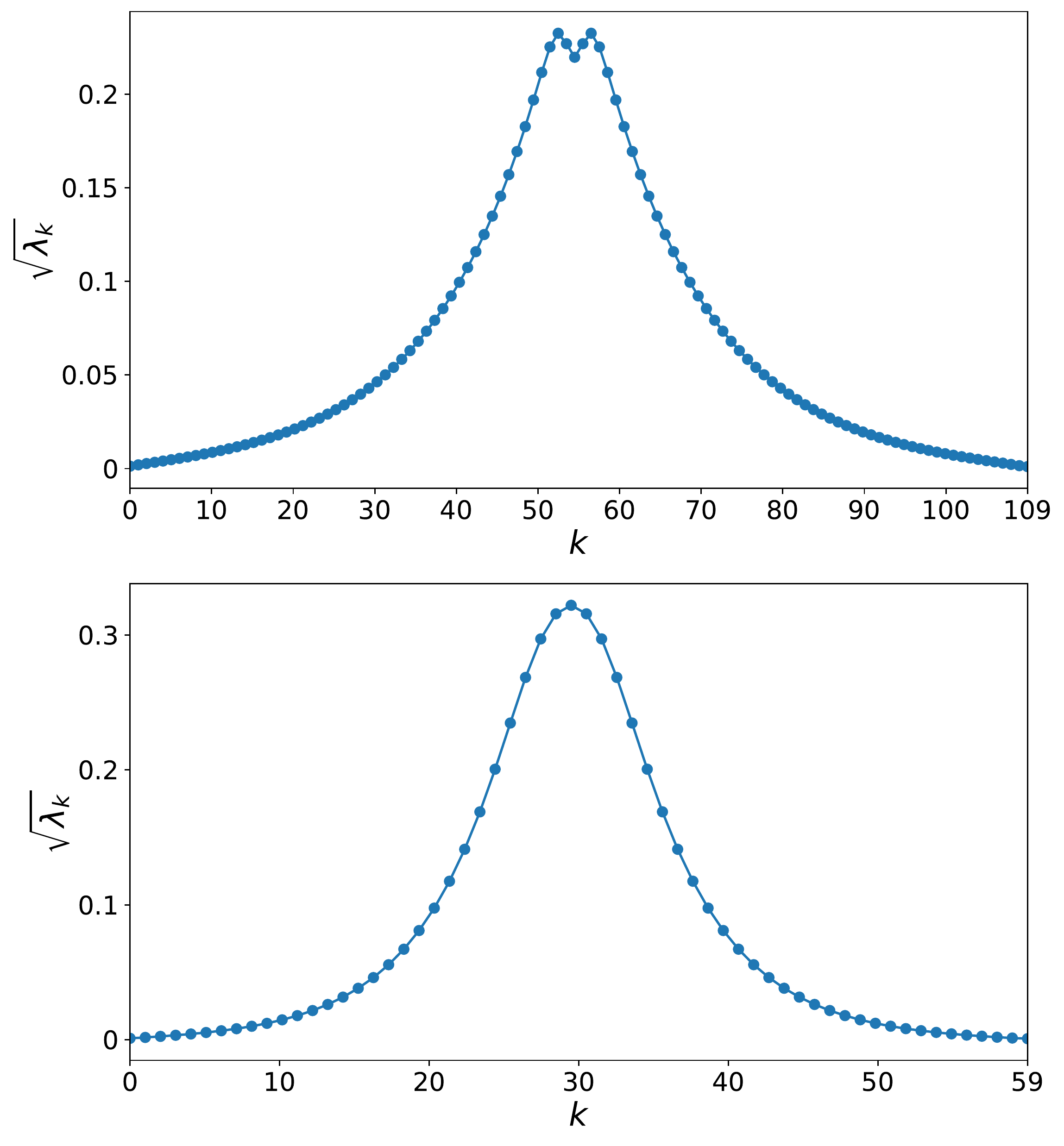}
\caption{Schmidt coefficients of the optimal state $\ket{\Psi^{(n)}}$ for the $(\alpha_1=0.75, \alpha_3=1.3)$ functional (top) and for the $(\alpha_1=0.75, \alpha_3=1.85)$ functional (bottom).}\label{FIG:10}
\end{figure}

\begin{figure}[t]
\includegraphics[width=8.5cm,trim=1cm 0cm 0cm 0cm;clip]{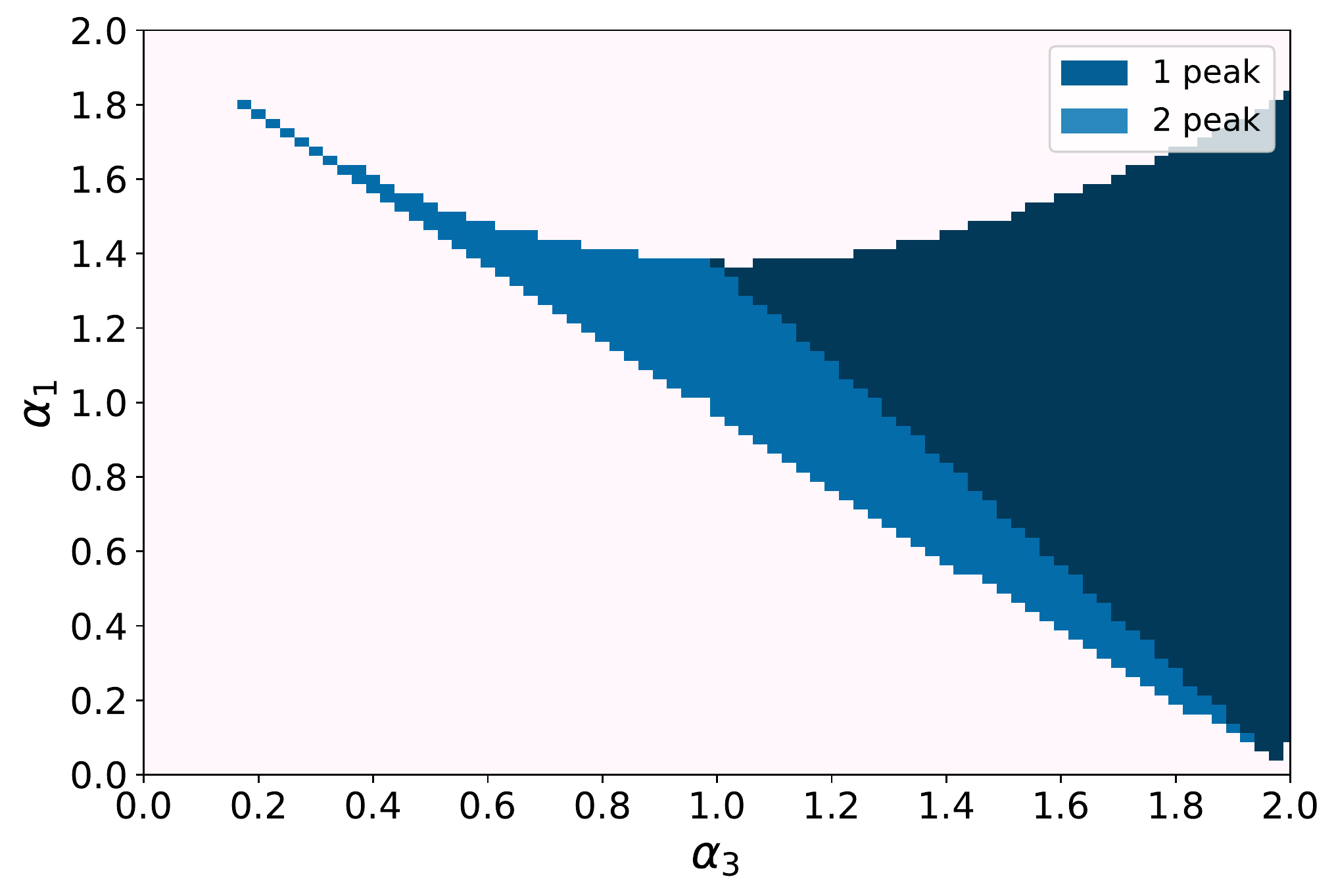}
\caption{Classification of solutions by their optimal state $\ket{\Psi^{(n)}}$. The labelled areas show the subregions in which the Schmidt coefficients exhibit a single maximum at the middle point or two maxima placed symmetrically about the middle point. As in FIG. \ref{FIG:8}, the plot shows only the functionals being classified.}\label{FIG:11}
\end{figure}

\begin{figure}[h]
\includegraphics[width=9.25cm,trim=0cm 0.75cm 0cm 0cm;clip]{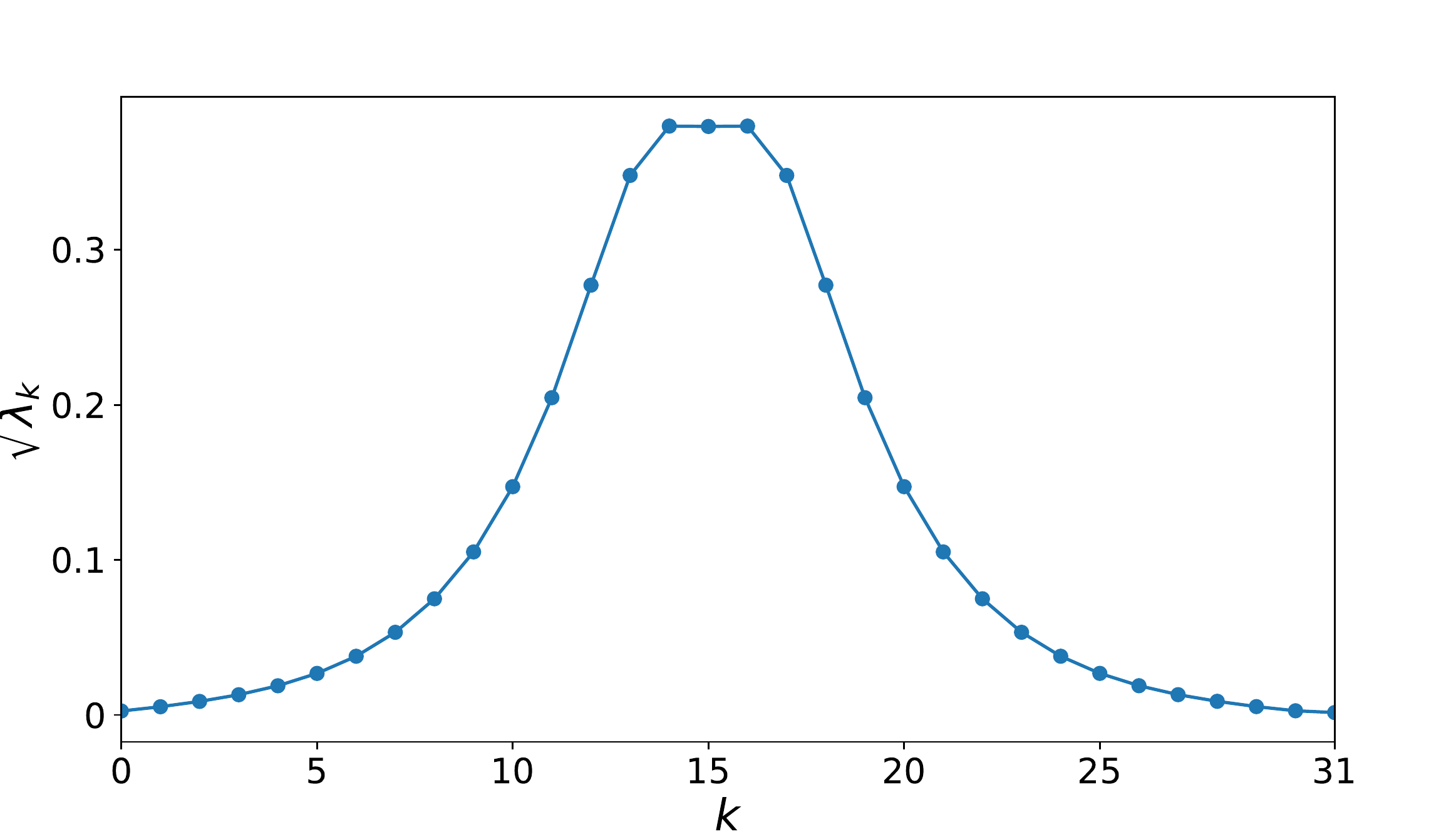}
\caption{Curve of Schmidt coefficients of the optimal state $\ket{\Psi^{(n)}}$ for the functional in $(\alpha_1=1, \alpha_3=1.275)$, a point on the boundary between the two regions depicted in FIG. \ref{FIG:11}. }\label{FIG:12}
\end{figure}

For higher dimensional solutions we can also look at the Schmidt coefficients in the optimal state in Eq.~\eqref{eq:pvst}. In FIG. \ref{FIG:10} we show these coefficients in the solutions for $(\alpha_1=0.75, \alpha_3=1.3)$ and $(\alpha_1=0.75, \alpha_3=1.85)$. Leaving aside the dimension of these solutions the most noticeable difference between the curves is seen in the vicinity of $\frac{n}{2}$, with the $\alpha_3=1.3$ curve showing two sharp peaks while the $\alpha_3=1.85$ curve shows one rounded peak. In fact, as $\alpha_3$ increases a gradual transition from the curve at the top of the figure to that at the bottom is observed, and this behaviour is reproduced for different values of $\alpha_1$. In FIG. \ref{FIG:11} we show the regions in which there is one or two maximal Schmidt coefficients. At the boundary between these regions, where the transition occurs, the maximal value is shared by three Schmidt coefficients and the middle part of the curve becomes flat, as shown in FIG. \ref{FIG:12}.

\begin{figure}[t]
\includegraphics[width=8.7cm,trim=0cm 0cm 0cm -0.7cm;clip]{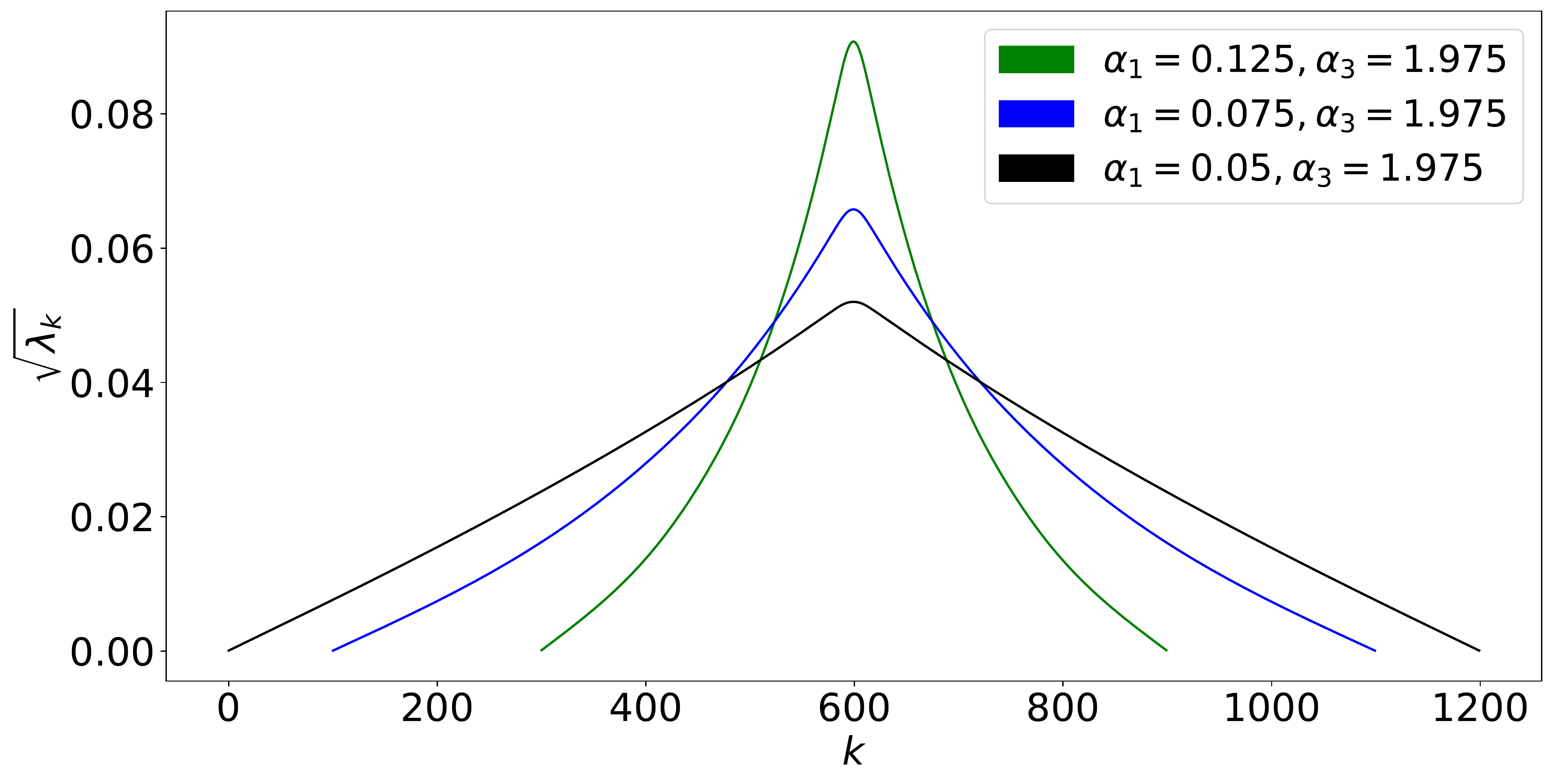}
\caption{Curve of Schmidt coefficients of the optimal state $\ket{\Psi^{(n)}}$ for the functionals in $(\alpha_1=0.05, \alpha_3=1.975)$, $(\alpha_1=0.075, \alpha_3=1.975)$ and $(\alpha_1=0.125, \alpha_3=1.975)$.}\label{FIG:13}
\end{figure}

Finally, in FIG.~\ref{FIG:13} we show the Schmidt coefficients of the optimal state for three different functionals, defined by parameters $(\alpha_1=0.05, \alpha_3=1.975)$, $(\alpha_1=0.075, \alpha_3=1.975)$ and $(\alpha_1=0.125, \alpha_3=1.975)$. Note that the tail of the curve straightens as $\alpha_1$ approaches $0$, a behaviour we consistently observe in the optimal solutions for the functionals near the $\alpha_1=-\alpha_3+2$ line. This behaviour is also observed when analysing, for fixed values of $\alpha_1, \alpha_3$, optimal states in dimensions lower than the needed to close the gap with the NPA upper bound. We therefore associate this feature with a slower convergence of the sequence $\{\beta_{PV}^{(n)}\}$ to the optimal value when the functionals approach the region in which $\beta_Q=4+\alpha_3^2$.

It is worth mentioning that we have also tried to classify the tails of the Schmidt curves by fitting them to decaying exponential or Gaussian functions, but this has not led to any conclusive results. Under the assumption that the almost constant value $\theta_c$ in FIG.~\ref{FIG:9} is exactly constant an analytic expression for this tail is easily derived, and we found this curve to fit the Schmidt coefficients of the optimal state better than the ones mentioned above in all high-dimensional solutions.

\section{Conclusions}

In this work we have considered a three-parameter family of functionals that extend those studied in Ref.~\cite{Ka.20} by adding a marginal contribution, while preserving their symmetries. This new family of functionals naturally splits into two branches that, although similar in appearance, exhibit quite different behaviours when it comes to the quantum value and the optimal realisations.

For the first branch we have shown that all functionals within the quantum advantage region, in the parameter space, admit a simple sum of squares decomposition which provides a tight upper bound to the quantum value. We made use of these SOS decompositions to characterise the set of quantum realisations providing the maximal value. Our results show that any partially entangled two-qubit state can be self-tested by maximally violating a given inequality in this branch, with the amount of entanglement being a function of the parameters defining the functional. The measurements providing the optimal value are not unique, but form a simple one-parameter family. These weak self-testing statements are easily seen to reduce to those made in Ref.~\cite{Ka.20} for the functionals in the parent family.

For the second family the analysis we provide is mainly numerical. Our results suggest that a large subset of these functionals attain their quantum value with a two-qubit realisation, which in some cases includes a trivial measurement by one of the parties. Outside this subset we tested the performance of a set of realisations known to attain the optimal value of the $I_{3322}$ functional \cite{PV.10}, which is a member of this family. In most of these functionals this particular strategy numerically attains the largest possible quantum value at a local dimension that depends on the specific value of the parameters defining the functional. Our results show how the coefficients parametrizing the optimal measurements and state change as the parameters defining the functionals vary. Our results suggest that the success of these particular realisations in maximally violating the $I_{3322}$ inequality is not related to $I_{3322}$ being a tight Bell inequality, but rather to the symmetries it exhibits.

\newpage

\section*{Acknowledgments}

We acknowledge support from the National Science Centre, Poland, under the SONATA project ``Fundamental aspects of the quantum set of
correlations'', Grant No.~2019/35/D/ST2/02014.

\bibliography{bibtex.bib}

\begin{thebibliography}{44}%
\makeatletter
\providecommand \@ifxundefined [1]{%
 \@ifx{#1\undefined}
}%
\providecommand \@ifnum [1]{%
 \ifnum #1\expandafter \@firstoftwo
 \else \expandafter \@secondoftwo
 \fi
}%
\providecommand \@ifx [1]{%
 \ifx #1\expandafter \@firstoftwo
 \else \expandafter \@secondoftwo
 \fi
}%
\providecommand \natexlab [1]{#1}%
\providecommand \enquote  [1]{``#1''}%
\providecommand \bibnamefont  [1]{#1}%
\providecommand \bibfnamefont [1]{#1}%
\providecommand \citenamefont [1]{#1}%
\providecommand \href@noop [0]{\@secondoftwo}%
\providecommand \href [0]{\begingroup \@sanitize@url \@href}%
\providecommand \@href[1]{\@@startlink{#1}\@@href}%
\providecommand \@@href[1]{\endgroup#1\@@endlink}%
\providecommand \@sanitize@url [0]{\catcode `\\12\catcode `\$12\catcode
  `\&12\catcode `\#12\catcode `\^12\catcode `\_12\catcode `\%12\relax}%
\providecommand \@@startlink[1]{}%
\providecommand \@@endlink[0]{}%
\providecommand \url  [0]{\begingroup\@sanitize@url \@url }%
\providecommand \@url [1]{\endgroup\@href {#1}{\urlprefix }}%
\providecommand \urlprefix  [0]{URL }%
\providecommand \Eprint [0]{\href }%
\providecommand \doibase [0]{http://dx.doi.org/}%
\providecommand \selectlanguage [0]{\@gobble}%
\providecommand \bibinfo  [0]{\@secondoftwo}%
\providecommand \bibfield  [0]{\@secondoftwo}%
\providecommand \translation [1]{[#1]}%
\providecommand \BibitemOpen [0]{}%
\providecommand \bibitemStop [0]{}%
\providecommand \bibitemNoStop [0]{.\EOS\space}%
\providecommand \EOS [0]{\spacefactor3000\relax}%
\providecommand \BibitemShut  [1]{\csname bibitem#1\endcsname}%
\let\auto@bib@innerbib\@empty
%</preamble>
\bibitem [{\citenamefont {Bell}(1964)}]{Be.64}%
  \BibitemOpen
  \bibfield  {author} {\bibinfo {author} {\bibfnamefont {J.~S.}\ \bibnamefont
  {Bell}},\ }\href {\doibase 10.1103/PhysicsPhysiqueFizika.1.195} {\bibfield
  {journal} {\bibinfo  {journal} {Physics Physique Fizika}\ }\textbf {\bibinfo
  {volume} {1}},\ \bibinfo {pages} {195} (\bibinfo {year} {1964})}\BibitemShut
  {NoStop}%
\bibitem [{\citenamefont {Brunner}\ \emph {et~al.}(2014)\citenamefont
  {Brunner}, \citenamefont {Cavalcanti}, \citenamefont {Pironio}, \citenamefont
  {Scarani},\ and\ \citenamefont {Wehner}}]{BC.14}%
  \BibitemOpen
  \bibfield  {author} {\bibinfo {author} {\bibfnamefont {N.}~\bibnamefont
  {Brunner}}, \bibinfo {author} {\bibfnamefont {D.}~\bibnamefont {Cavalcanti}},
  \bibinfo {author} {\bibfnamefont {S.}~\bibnamefont {Pironio}}, \bibinfo
  {author} {\bibfnamefont {V.}~\bibnamefont {Scarani}}, \ and\ \bibinfo
  {author} {\bibfnamefont {S.}~\bibnamefont {Wehner}},\ }\href {\doibase
  10.1103/RevModPhys.86.419} {\bibfield  {journal} {\bibinfo  {journal} {Rev.
  Mod. Phys.}\ }\textbf {\bibinfo {volume} {86}},\ \bibinfo {pages} {419}
  (\bibinfo {year} {2014})}\BibitemShut {NoStop}%
\bibitem [{\citenamefont {Barrett}\ \emph {et~al.}(2005)\citenamefont
  {Barrett}, \citenamefont {Hardy},\ and\ \citenamefont {Kent}}]{BH.95}%
  \BibitemOpen
  \bibfield  {author} {\bibinfo {author} {\bibfnamefont {J.}~\bibnamefont
  {Barrett}}, \bibinfo {author} {\bibfnamefont {L.}~\bibnamefont {Hardy}}, \
  and\ \bibinfo {author} {\bibfnamefont {A.}~\bibnamefont {Kent}},\ }\href
  {\doibase 10.1103/PhysRevLett.95.010503} {\bibfield  {journal} {\bibinfo
  {journal} {Phys. Rev. Lett.}\ }\textbf {\bibinfo {volume} {95}},\ \bibinfo
  {pages} {010503} (\bibinfo {year} {2005})}\BibitemShut {NoStop}%
\bibitem [{\citenamefont {Ac\'{\i}n}\ \emph
  {et~al.}(2006{\natexlab{a}})\citenamefont {Ac\'{\i}n}, \citenamefont
  {Gisin},\ and\ \citenamefont {Masanes}}]{AG.97}%
  \BibitemOpen
  \bibfield  {author} {\bibinfo {author} {\bibfnamefont {A.}~\bibnamefont
  {Ac\'{\i}n}}, \bibinfo {author} {\bibfnamefont {N.}~\bibnamefont {Gisin}}, \
  and\ \bibinfo {author} {\bibfnamefont {L.}~\bibnamefont {Masanes}},\ }\href
  {\doibase 10.1103/PhysRevLett.97.120405} {\bibfield  {journal} {\bibinfo
  {journal} {Phys. Rev. Lett.}\ }\textbf {\bibinfo {volume} {97}},\ \bibinfo
  {pages} {120405} (\bibinfo {year} {2006}{\natexlab{a}})}\BibitemShut
  {NoStop}%
\bibitem [{\citenamefont {Ac\'{\i}n}\ \emph
  {et~al.}(2006{\natexlab{b}})\citenamefont {Ac\'{\i}n}, \citenamefont
  {Gisin},\ and\ \citenamefont {Masanes}}]{GM.97}%
  \BibitemOpen
  \bibfield  {author} {\bibinfo {author} {\bibfnamefont {A.}~\bibnamefont
  {Ac\'{\i}n}}, \bibinfo {author} {\bibfnamefont {N.}~\bibnamefont {Gisin}}, \
  and\ \bibinfo {author} {\bibfnamefont {L.}~\bibnamefont {Masanes}},\ }\href
  {\doibase 10.1103/PhysRevLett.97.120405} {\bibfield  {journal} {\bibinfo
  {journal} {Phys. Rev. Lett.}\ }\textbf {\bibinfo {volume} {97}},\ \bibinfo
  {pages} {120405} (\bibinfo {year} {2006}{\natexlab{b}})}\BibitemShut
  {NoStop}%
\bibitem [{\citenamefont {Colbeck}(2011)}]{Co.11}%
  \BibitemOpen
  \bibfield  {author} {\bibinfo {author} {\bibfnamefont {R.}~\bibnamefont
  {Colbeck}},\ }\href@noop {} {\enquote {\bibinfo {title} {Quantum and
  relativistic protocols for secure multi-party computation},}\ } (\bibinfo
  {year} {2011}),\ \Eprint {http://arxiv.org/abs/0911.3814} {arXiv:0911.3814
  [quant-ph]} \BibitemShut {NoStop}%
\bibitem [{\citenamefont {Ac\'{\i}n}\ \emph {et~al.}(2007)\citenamefont
  {Ac\'{\i}n}, \citenamefont {Brunner}, \citenamefont {Gisin}, \citenamefont
  {Massar}, \citenamefont {Pironio},\ and\ \citenamefont {Scarani}}]{AB.98}%
  \BibitemOpen
  \bibfield  {author} {\bibinfo {author} {\bibfnamefont {A.}~\bibnamefont
  {Ac\'{\i}n}}, \bibinfo {author} {\bibfnamefont {N.}~\bibnamefont {Brunner}},
  \bibinfo {author} {\bibfnamefont {N.}~\bibnamefont {Gisin}}, \bibinfo
  {author} {\bibfnamefont {S.}~\bibnamefont {Massar}}, \bibinfo {author}
  {\bibfnamefont {S.}~\bibnamefont {Pironio}}, \ and\ \bibinfo {author}
  {\bibfnamefont {V.}~\bibnamefont {Scarani}},\ }\href {\doibase
  10.1103/PhysRevLett.98.230501} {\bibfield  {journal} {\bibinfo  {journal}
  {Phys. Rev. Lett.}\ }\textbf {\bibinfo {volume} {98}},\ \bibinfo {pages}
  {230501} (\bibinfo {year} {2007})}\BibitemShut {NoStop}%
\bibitem [{\citenamefont {Pironio}\ \emph {et~al.}(2010)\citenamefont
  {Pironio}, \citenamefont {Ac{\'i}n}, \citenamefont {Massar}, \citenamefont
  {de~la Giroday}, \citenamefont {Matsukevich}, \citenamefont {Maunz},
  \citenamefont {Olmschenk}, \citenamefont {Hayes}, \citenamefont {Luo},
  \citenamefont {Manning},\ and\ \citenamefont {Monroe}}]{Pi.10}%
  \BibitemOpen
  \bibfield  {author} {\bibinfo {author} {\bibfnamefont {S.}~\bibnamefont
  {Pironio}}, \bibinfo {author} {\bibfnamefont {A.}~\bibnamefont {Ac{\'i}n}},
  \bibinfo {author} {\bibfnamefont {S.}~\bibnamefont {Massar}}, \bibinfo
  {author} {\bibfnamefont {A.~B.}\ \bibnamefont {de~la Giroday}}, \bibinfo
  {author} {\bibfnamefont {D.~N.}\ \bibnamefont {Matsukevich}}, \bibinfo
  {author} {\bibfnamefont {P.}~\bibnamefont {Maunz}}, \bibinfo {author}
  {\bibfnamefont {S.}~\bibnamefont {Olmschenk}}, \bibinfo {author}
  {\bibfnamefont {D.}~\bibnamefont {Hayes}}, \bibinfo {author} {\bibfnamefont
  {L.}~\bibnamefont {Luo}}, \bibinfo {author} {\bibfnamefont {T.~A.}\
  \bibnamefont {Manning}}, \ and\ \bibinfo {author} {\bibfnamefont
  {C.}~\bibnamefont {Monroe}},\ }\href {\doibase 10.1038/nature09008}
  {\bibfield  {journal} {\bibinfo  {journal} {Nature}\ }\textbf {\bibinfo
  {volume} {464}},\ \bibinfo {pages} {1021} (\bibinfo {year}
  {2010})}\BibitemShut {NoStop}%
\bibitem [{\citenamefont {Ekert}\ and\ \citenamefont {Renner}(2014)}]{Ek.14}%
  \BibitemOpen
  \bibfield  {author} {\bibinfo {author} {\bibfnamefont {A.}~\bibnamefont
  {Ekert}}\ and\ \bibinfo {author} {\bibfnamefont {R.}~\bibnamefont {Renner}},\
  }\href {\doibase 10.1038/nature13132} {\bibfield  {journal} {\bibinfo
  {journal} {Nature}\ }\textbf {\bibinfo {volume} {507}},\ \bibinfo {pages}
  {443} (\bibinfo {year} {2014})}\BibitemShut {NoStop}%
\bibitem [{\citenamefont {Pirandola}\ \emph {et~al.}(2020)\citenamefont
  {Pirandola}, \citenamefont {Andersen}, \citenamefont {Banchi}, \citenamefont
  {Berta}, \citenamefont {Bunandar}, \citenamefont {Colbeck}, \citenamefont
  {Englund}, \citenamefont {Gehring}, \citenamefont {Lupo}, \citenamefont
  {Ottaviani}, \citenamefont {Pereira}, \citenamefont {Razavi}, \citenamefont
  {Shaari}, \citenamefont {Tomamichel}, \citenamefont {Usenko}, \citenamefont
  {Vallone}, \citenamefont {Villoresi},\ and\ \citenamefont {Wallden}}]{Pi.20}%
  \BibitemOpen
  \bibfield  {author} {\bibinfo {author} {\bibfnamefont {S.}~\bibnamefont
  {Pirandola}}, \bibinfo {author} {\bibfnamefont {U.~L.}\ \bibnamefont
  {Andersen}}, \bibinfo {author} {\bibfnamefont {L.}~\bibnamefont {Banchi}},
  \bibinfo {author} {\bibfnamefont {M.}~\bibnamefont {Berta}}, \bibinfo
  {author} {\bibfnamefont {D.}~\bibnamefont {Bunandar}}, \bibinfo {author}
  {\bibfnamefont {R.}~\bibnamefont {Colbeck}}, \bibinfo {author} {\bibfnamefont
  {D.}~\bibnamefont {Englund}}, \bibinfo {author} {\bibfnamefont
  {T.}~\bibnamefont {Gehring}}, \bibinfo {author} {\bibfnamefont
  {C.}~\bibnamefont {Lupo}}, \bibinfo {author} {\bibfnamefont {C.}~\bibnamefont
  {Ottaviani}}, \bibinfo {author} {\bibfnamefont {J.~L.}\ \bibnamefont
  {Pereira}}, \bibinfo {author} {\bibfnamefont {M.}~\bibnamefont {Razavi}},
  \bibinfo {author} {\bibfnamefont {J.~S.}\ \bibnamefont {Shaari}}, \bibinfo
  {author} {\bibfnamefont {M.}~\bibnamefont {Tomamichel}}, \bibinfo {author}
  {\bibfnamefont {V.~C.}\ \bibnamefont {Usenko}}, \bibinfo {author}
  {\bibfnamefont {G.}~\bibnamefont {Vallone}}, \bibinfo {author} {\bibfnamefont
  {P.}~\bibnamefont {Villoresi}}, \ and\ \bibinfo {author} {\bibfnamefont
  {P.}~\bibnamefont {Wallden}},\ }\href {\doibase 10.1364/AOP.361502}
  {\bibfield  {journal} {\bibinfo  {journal} {Adv. Opt. Photon.}\ }\textbf
  {\bibinfo {volume} {12}},\ \bibinfo {pages} {1012} (\bibinfo {year}
  {2020})}\BibitemShut {NoStop}%
\bibitem [{\citenamefont {Ac{\'i}n}\ and\ \citenamefont
  {Masanes}(2016)}]{AM.16}%
  \BibitemOpen
  \bibfield  {author} {\bibinfo {author} {\bibfnamefont {A.}~\bibnamefont
  {Ac{\'i}n}}\ and\ \bibinfo {author} {\bibfnamefont {L.}~\bibnamefont
  {Masanes}},\ }\href {\doibase 10.1038/nature20119} {\bibfield  {journal}
  {\bibinfo  {journal} {Nature}\ }\textbf {\bibinfo {volume} {540}},\ \bibinfo
  {pages} {213} (\bibinfo {year} {2016})}\BibitemShut {NoStop}%
\bibitem [{\citenamefont {Bera}\ \emph {et~al.}(2017)\citenamefont {Bera},
  \citenamefont {Ac{\'{\i}}n}, \citenamefont {Ku{\'{s}}}, \citenamefont
  {Mitchell},\ and\ \citenamefont {Lewenstein}}]{BA.17}%
  \BibitemOpen
  \bibfield  {author} {\bibinfo {author} {\bibfnamefont {M.~N.}\ \bibnamefont
  {Bera}}, \bibinfo {author} {\bibfnamefont {A.}~\bibnamefont {Ac{\'{\i}}n}},
  \bibinfo {author} {\bibfnamefont {M.}~\bibnamefont {Ku{\'{s}}}}, \bibinfo
  {author} {\bibfnamefont {M.~W.}\ \bibnamefont {Mitchell}}, \ and\ \bibinfo
  {author} {\bibfnamefont {M.}~\bibnamefont {Lewenstein}},\ }\href {\doibase
  10.1088/1361-6633/aa8731} {\bibfield  {journal} {\bibinfo  {journal} {Reports
  on Progress in Physics}\ }\textbf {\bibinfo {volume} {80}},\ \bibinfo {pages}
  {124001} (\bibinfo {year} {2017})}\BibitemShut {NoStop}%
\bibitem [{\citenamefont {Brunner}(2014)}]{Br.14}%
  \BibitemOpen
  \bibfield  {author} {\bibinfo {author} {\bibfnamefont {N.}~\bibnamefont
  {Brunner}},\ }in\ \href {\doibase 10.1364/QIM.2014.QW3A.2} {\emph {\bibinfo
  {booktitle} {Research in Optical Sciences}}}\ (\bibinfo  {publisher} {Optical
  Society of America},\ \bibinfo {year} {2014})\ p.\ \bibinfo {pages}
  {QW3A.2}\BibitemShut {NoStop}%
\bibitem [{\citenamefont {Tsirel'son}(1987)}]{Ts.87}%
  \BibitemOpen
  \bibfield  {author} {\bibinfo {author} {\bibfnamefont {B.~S.}\ \bibnamefont
  {Tsirel'son}},\ }\href {\doibase 10.1007/BF01663472} {\bibfield  {journal}
  {\bibinfo  {journal} {Journal of Soviet Mathematics}\ }\textbf {\bibinfo
  {volume} {36}},\ \bibinfo {pages} {557} (\bibinfo {year} {1987})}\BibitemShut
  {NoStop}%
\bibitem [{\citenamefont {Tsirel'son}(1993)}]{Ts.93}%
  \BibitemOpen
  \bibfield  {author} {\bibinfo {author} {\bibfnamefont {B.~S.}\ \bibnamefont
  {Tsirel'son}},\ }\href {https://ci.nii.ac.jp/naid/10026857475/en/} {\bibfield
   {journal} {\bibinfo  {journal} {Hadronic Journal Supplement}\ }\textbf
  {\bibinfo {volume} {8}},\ \bibinfo {pages} {329} (\bibinfo {year}
  {1993})}\BibitemShut {NoStop}%
\bibitem [{\citenamefont {Summers}\ and\ \citenamefont {Werner}(1987)}]{Su.87}%
  \BibitemOpen
  \bibfield  {author} {\bibinfo {author} {\bibfnamefont {S.~J.}\ \bibnamefont
  {Summers}}\ and\ \bibinfo {author} {\bibfnamefont {R.}~\bibnamefont
  {Werner}},\ }\href {\doibase 10.1007/BF01207366} {\bibfield  {journal}
  {\bibinfo  {journal} {Communications in Mathematical Physics}\ }\textbf
  {\bibinfo {volume} {110}},\ \bibinfo {pages} {247} (\bibinfo {year}
  {1987})}\BibitemShut {NoStop}%
\bibitem [{\citenamefont {Popescu}\ and\ \citenamefont
  {Rohrlich}(1992)}]{PR.92}%
  \BibitemOpen
  \bibfield  {author} {\bibinfo {author} {\bibfnamefont {S.}~\bibnamefont
  {Popescu}}\ and\ \bibinfo {author} {\bibfnamefont {D.}~\bibnamefont
  {Rohrlich}},\ }\href {\doibase https://doi.org/10.1016/0375-9601(92)90819-8}
  {\bibfield  {journal} {\bibinfo  {journal} {Physics Letters A}\ }\textbf
  {\bibinfo {volume} {169}},\ \bibinfo {pages} {411} (\bibinfo {year}
  {1992})}\BibitemShut {NoStop}%
\bibitem [{\citenamefont {{\v{S}}upi{\'{c}}}\ and\ \citenamefont
  {Bowles}(2020)}]{Su.20}%
  \BibitemOpen
  \bibfield  {author} {\bibinfo {author} {\bibfnamefont {I.}~\bibnamefont
  {{\v{S}}upi{\'{c}}}}\ and\ \bibinfo {author} {\bibfnamefont {J.}~\bibnamefont
  {Bowles}},\ }\href {\doibase 10.22331/q-2020-09-30-337} {\bibfield  {journal}
  {\bibinfo  {journal} {{Quantum}}\ }\textbf {\bibinfo {volume} {4}},\ \bibinfo
  {pages} {337} (\bibinfo {year} {2020})}\BibitemShut {NoStop}%
\bibitem [{\citenamefont {Mayers}\ and\ \citenamefont {Yao}(2004)}]{MY.04}%
  \BibitemOpen
  \bibfield  {author} {\bibinfo {author} {\bibfnamefont {D.}~\bibnamefont
  {Mayers}}\ and\ \bibinfo {author} {\bibfnamefont {A.}~\bibnamefont {Yao}},\
  }\href@noop {} {\enquote {\bibinfo {title} {Self testing quantum
  apparatus},}\ } (\bibinfo {year} {2004}),\ \Eprint
  {http://arxiv.org/abs/quant-ph/0307205} {arXiv:quant-ph/0307205 [quant-ph]}
  \BibitemShut {NoStop}%
\bibitem [{\citenamefont {Mayers}\ and\ \citenamefont {Yao}(1998)}]{MY.98}%
  \BibitemOpen
  \bibfield  {author} {\bibinfo {author} {\bibfnamefont {D.}~\bibnamefont
  {Mayers}}\ and\ \bibinfo {author} {\bibfnamefont {A.}~\bibnamefont {Yao}},\
  }\href@noop {} {\enquote {\bibinfo {title} {Quantum cryptography with
  imperfect apparatus},}\ } (\bibinfo {year} {1998}),\ \Eprint
  {http://arxiv.org/abs/quant-ph/9809039} {arXiv:quant-ph/9809039 [quant-ph]}
  \BibitemShut {NoStop}%
\bibitem [{\citenamefont {Reichardt}\ \emph {et~al.}(2013)\citenamefont
  {Reichardt}, \citenamefont {Unger},\ and\ \citenamefont {Vazirani}}]{RU.13}%
  \BibitemOpen
  \bibfield  {author} {\bibinfo {author} {\bibfnamefont {B.~W.}\ \bibnamefont
  {Reichardt}}, \bibinfo {author} {\bibfnamefont {F.}~\bibnamefont {Unger}}, \
  and\ \bibinfo {author} {\bibfnamefont {U.}~\bibnamefont {Vazirani}},\ }\href
  {\doibase 10.1038/nature12035} {\bibfield  {journal} {\bibinfo  {journal}
  {Nature}\ }\textbf {\bibinfo {volume} {496}},\ \bibinfo {pages} {456}
  (\bibinfo {year} {2013})}\BibitemShut {NoStop}%
\bibitem [{\citenamefont {Goh}\ \emph {et~al.}(2018)\citenamefont {Goh},
  \citenamefont {Kaniewski}, \citenamefont {Wolfe}, \citenamefont {V\'ertesi},
  \citenamefont {Wu}, \citenamefont {Cai}, \citenamefont {Liang},\ and\
  \citenamefont {Scarani}}]{GK.18}%
  \BibitemOpen
  \bibfield  {author} {\bibinfo {author} {\bibfnamefont {K.~T.}\ \bibnamefont
  {Goh}}, \bibinfo {author} {\bibfnamefont {J.}~\bibnamefont {Kaniewski}},
  \bibinfo {author} {\bibfnamefont {E.}~\bibnamefont {Wolfe}}, \bibinfo
  {author} {\bibfnamefont {T.}~\bibnamefont {V\'ertesi}}, \bibinfo {author}
  {\bibfnamefont {X.}~\bibnamefont {Wu}}, \bibinfo {author} {\bibfnamefont
  {Y.}~\bibnamefont {Cai}}, \bibinfo {author} {\bibfnamefont {Y.-C.}\
  \bibnamefont {Liang}}, \ and\ \bibinfo {author} {\bibfnamefont
  {V.}~\bibnamefont {Scarani}},\ }\href {\doibase 10.1103/PhysRevA.97.022104}
  {\bibfield  {journal} {\bibinfo  {journal} {Phys. Rev. A}\ }\textbf {\bibinfo
  {volume} {97}},\ \bibinfo {pages} {022104} (\bibinfo {year}
  {2018})}\BibitemShut {NoStop}%
\bibitem [{\citenamefont {Salavrakos}\ \emph {et~al.}(2017)\citenamefont
  {Salavrakos}, \citenamefont {Augusiak}, \citenamefont {Tura}, \citenamefont
  {Wittek}, \citenamefont {Ac\'{\i}n},\ and\ \citenamefont {Pironio}}]{SA.17}%
  \BibitemOpen
  \bibfield  {author} {\bibinfo {author} {\bibfnamefont {A.}~\bibnamefont
  {Salavrakos}}, \bibinfo {author} {\bibfnamefont {R.}~\bibnamefont
  {Augusiak}}, \bibinfo {author} {\bibfnamefont {J.}~\bibnamefont {Tura}},
  \bibinfo {author} {\bibfnamefont {P.}~\bibnamefont {Wittek}}, \bibinfo
  {author} {\bibfnamefont {A.}~\bibnamefont {Ac\'{\i}n}}, \ and\ \bibinfo
  {author} {\bibfnamefont {S.}~\bibnamefont {Pironio}},\ }\href {\doibase
  10.1103/PhysRevLett.119.040402} {\bibfield  {journal} {\bibinfo  {journal}
  {Phys. Rev. Lett.}\ }\textbf {\bibinfo {volume} {119}},\ \bibinfo {pages}
  {040402} (\bibinfo {year} {2017})}\BibitemShut {NoStop}%
\bibitem [{\citenamefont {Baccari}\ \emph {et~al.}(2020)\citenamefont
  {Baccari}, \citenamefont {Augusiak}, \citenamefont {\ifmmode \check{S}\else
  \v{S}\fi{}upi\ifmmode~\acute{c}\else \'{c}\fi{}}, \citenamefont {Tura},\ and\
  \citenamefont {Ac\'{\i}n}}]{BA.20}%
  \BibitemOpen
  \bibfield  {author} {\bibinfo {author} {\bibfnamefont {F.}~\bibnamefont
  {Baccari}}, \bibinfo {author} {\bibfnamefont {R.}~\bibnamefont {Augusiak}},
  \bibinfo {author} {\bibfnamefont {I.}~\bibnamefont {\ifmmode \check{S}\else
  \v{S}\fi{}upi\ifmmode~\acute{c}\else \'{c}\fi{}}}, \bibinfo {author}
  {\bibfnamefont {J.}~\bibnamefont {Tura}}, \ and\ \bibinfo {author}
  {\bibfnamefont {A.}~\bibnamefont {Ac\'{\i}n}},\ }\href {\doibase
  10.1103/PhysRevLett.124.020402} {\bibfield  {journal} {\bibinfo  {journal}
  {Phys. Rev. Lett.}\ }\textbf {\bibinfo {volume} {124}},\ \bibinfo {pages}
  {020402} (\bibinfo {year} {2020})}\BibitemShut {NoStop}%
\bibitem [{\citenamefont {Bamps}\ and\ \citenamefont {Pironio}(2015)}]{BP.15}%
  \BibitemOpen
  \bibfield  {author} {\bibinfo {author} {\bibfnamefont {C.}~\bibnamefont
  {Bamps}}\ and\ \bibinfo {author} {\bibfnamefont {S.}~\bibnamefont
  {Pironio}},\ }\href {\doibase 10.1103/PhysRevA.91.052111} {\bibfield
  {journal} {\bibinfo  {journal} {Phys. Rev. A}\ }\textbf {\bibinfo {volume}
  {91}},\ \bibinfo {pages} {052111} (\bibinfo {year} {2015})}\BibitemShut
  {NoStop}%
\bibitem [{\citenamefont {{\v{S}}upi{\'{c}}}\ \emph {et~al.}(2016)\citenamefont
  {{\v{S}}upi{\'{c}}}, \citenamefont {Augusiak}, \citenamefont {Salavrakos},\
  and\ \citenamefont {Ac{\'{\i}}n}}]{SA.16}%
  \BibitemOpen
  \bibfield  {author} {\bibinfo {author} {\bibfnamefont {I.}~\bibnamefont
  {{\v{S}}upi{\'{c}}}}, \bibinfo {author} {\bibfnamefont {R.}~\bibnamefont
  {Augusiak}}, \bibinfo {author} {\bibfnamefont {A.}~\bibnamefont
  {Salavrakos}}, \ and\ \bibinfo {author} {\bibfnamefont {A.}~\bibnamefont
  {Ac{\'{\i}}n}},\ }\href {\doibase 10.1088/1367-2630/18/3/035013} {\bibfield
  {journal} {\bibinfo  {journal} {New Journal of Physics}\ }\textbf {\bibinfo
  {volume} {18}},\ \bibinfo {pages} {035013} (\bibinfo {year}
  {2016})}\BibitemShut {NoStop}%
\bibitem [{\citenamefont {Kaniewski}(2016)}]{Ka.16}%
  \BibitemOpen
  \bibfield  {author} {\bibinfo {author} {\bibfnamefont {J.}~\bibnamefont
  {Kaniewski}},\ }\href {\doibase 10.1103/PhysRevLett.117.070402} {\bibfield
  {journal} {\bibinfo  {journal} {Phys. Rev. Lett.}\ }\textbf {\bibinfo
  {volume} {117}},\ \bibinfo {pages} {070402} (\bibinfo {year}
  {2016})}\BibitemShut {NoStop}%
\bibitem [{\citenamefont {Andersson}\ \emph {et~al.}(2017)\citenamefont
  {Andersson}, \citenamefont {Badzikag}, \citenamefont {Bengtsson},
  \citenamefont {Dumitru},\ and\ \citenamefont {Cabello}}]{AB.17}%
  \BibitemOpen
  \bibfield  {author} {\bibinfo {author} {\bibfnamefont {O.}~\bibnamefont
  {Andersson}}, \bibinfo {author} {\bibfnamefont {P.}~\bibnamefont {Badzikag}},
  \bibinfo {author} {\bibfnamefont {I.}~\bibnamefont {Bengtsson}}, \bibinfo
  {author} {\bibfnamefont {I.}~\bibnamefont {Dumitru}}, \ and\ \bibinfo
  {author} {\bibfnamefont {A.}~\bibnamefont {Cabello}},\ }\href {\doibase
  10.1103/PhysRevA.96.032119} {\bibfield  {journal} {\bibinfo  {journal} {Phys.
  Rev. A}\ }\textbf {\bibinfo {volume} {96}},\ \bibinfo {pages} {032119}
  (\bibinfo {year} {2017})}\BibitemShut {NoStop}%
\bibitem [{\citenamefont {Wagner}\ \emph {et~al.}(2020)\citenamefont {Wagner},
  \citenamefont {Bancal}, \citenamefont {Sangouard},\ and\ \citenamefont
  {Sekatski}}]{WB.20}%
  \BibitemOpen
  \bibfield  {author} {\bibinfo {author} {\bibfnamefont {S.}~\bibnamefont
  {Wagner}}, \bibinfo {author} {\bibfnamefont {J.-D.}\ \bibnamefont {Bancal}},
  \bibinfo {author} {\bibfnamefont {N.}~\bibnamefont {Sangouard}}, \ and\
  \bibinfo {author} {\bibfnamefont {P.}~\bibnamefont {Sekatski}},\ }\href
  {\doibase 10.22331/q-2020-03-19-243} {\bibfield  {journal} {\bibinfo
  {journal} {{Quantum}}\ }\textbf {\bibinfo {volume} {4}},\ \bibinfo {pages}
  {243} (\bibinfo {year} {2020})}\BibitemShut {NoStop}%
\bibitem [{\citenamefont {Tavakoli}\ \emph {et~al.}(2021)\citenamefont
  {Tavakoli}, \citenamefont {Farkas}, \citenamefont {Rosset}, \citenamefont
  {Bancal},\ and\ \citenamefont {Kaniewski}}]{TF.21}%
  \BibitemOpen
  \bibfield  {author} {\bibinfo {author} {\bibfnamefont {A.}~\bibnamefont
  {Tavakoli}}, \bibinfo {author} {\bibfnamefont {M.}~\bibnamefont {Farkas}},
  \bibinfo {author} {\bibfnamefont {D.}~\bibnamefont {Rosset}}, \bibinfo
  {author} {\bibfnamefont {J.-D.}\ \bibnamefont {Bancal}}, \ and\ \bibinfo
  {author} {\bibfnamefont {J.}~\bibnamefont {Kaniewski}},\ }\href {\doibase
  10.1126/sciadv.abc3847} {\bibfield  {journal} {\bibinfo  {journal} {Science
  Advances}\ }\textbf {\bibinfo {volume} {7}},\ \bibinfo {pages} {eabc3847}
  (\bibinfo {year} {2021})},\ \Eprint
  {http://arxiv.org/abs/https://www.science.org/doi/pdf/10.1126/sciadv.abc3847}
  {https://www.science.org/doi/pdf/10.1126/sciadv.abc3847} \BibitemShut
  {NoStop}%
\bibitem [{\citenamefont {Kaniewski}(2020)}]{Ka.20}%
  \BibitemOpen
  \bibfield  {author} {\bibinfo {author} {\bibfnamefont {J.}~\bibnamefont
  {Kaniewski}},\ }\href {\doibase 10.1103/PhysRevResearch.2.033420} {\bibfield
  {journal} {\bibinfo  {journal} {Phys. Rev. Research}\ }\textbf {\bibinfo
  {volume} {2}},\ \bibinfo {pages} {033420} (\bibinfo {year}
  {2020})}\BibitemShut {NoStop}%
\bibitem [{\citenamefont {Froissart}(1981)}]{Fr.81}%
  \BibitemOpen
  \bibfield  {author} {\bibinfo {author} {\bibfnamefont {M.}~\bibnamefont
  {Froissart}},\ }\href@noop {} {\bibfield  {journal} {\bibinfo  {journal} {Il
  Nuovo Cimento B (1971-1996)}\ }\textbf {\bibinfo {volume} {64}},\ \bibinfo
  {pages} {241} (\bibinfo {year} {1981})}\BibitemShut {NoStop}%
\bibitem [{\citenamefont {Śliwa}(2003)}]{Sl.03}%
  \BibitemOpen
  \bibfield  {author} {\bibinfo {author} {\bibfnamefont {C.}~\bibnamefont
  {Śliwa}},\ }\href {\doibase https://doi.org/10.1016/S0375-9601(03)01115-0}
  {\bibfield  {journal} {\bibinfo  {journal} {Physics Letters A}\ }\textbf
  {\bibinfo {volume} {317}},\ \bibinfo {pages} {165} (\bibinfo {year}
  {2003})}\BibitemShut {NoStop}%
\bibitem [{\citenamefont {Collins}\ and\ \citenamefont {Gisin}(2004)}]{CG.04}%
  \BibitemOpen
  \bibfield  {author} {\bibinfo {author} {\bibfnamefont {D.}~\bibnamefont
  {Collins}}\ and\ \bibinfo {author} {\bibfnamefont {N.}~\bibnamefont
  {Gisin}},\ }\href {\doibase 10.1088/0305-4470/37/5/021} {\bibfield  {journal}
  {\bibinfo  {journal} {Journal of Physics A: Mathematical and General}\
  }\textbf {\bibinfo {volume} {37}},\ \bibinfo {pages} {1775} (\bibinfo {year}
  {2004})}\BibitemShut {NoStop}%
\bibitem [{\citenamefont {P\'al}\ and\ \citenamefont
  {V\'ertesi}(2010)}]{PV.10}%
  \BibitemOpen
  \bibfield  {author} {\bibinfo {author} {\bibfnamefont {K.~F.}\ \bibnamefont
  {P\'al}}\ and\ \bibinfo {author} {\bibfnamefont {T.}~\bibnamefont
  {V\'ertesi}},\ }\href {\doibase 10.1103/PhysRevA.82.022116} {\bibfield
  {journal} {\bibinfo  {journal} {Phys. Rev. A}\ }\textbf {\bibinfo {volume}
  {82}},\ \bibinfo {pages} {022116} (\bibinfo {year} {2010})}\BibitemShut
  {NoStop}%
\bibitem [{\citenamefont {Barrett}\ and\ \citenamefont
  {Pironio}(2005)}]{BP.05}%
  \BibitemOpen
  \bibfield  {author} {\bibinfo {author} {\bibfnamefont {J.}~\bibnamefont
  {Barrett}}\ and\ \bibinfo {author} {\bibfnamefont {S.}~\bibnamefont
  {Pironio}},\ }\href {\doibase 10.1103/PhysRevLett.95.140401} {\bibfield
  {journal} {\bibinfo  {journal} {Phys. Rev. Lett.}\ }\textbf {\bibinfo
  {volume} {95}},\ \bibinfo {pages} {140401} (\bibinfo {year}
  {2005})}\BibitemShut {NoStop}%
\bibitem [{\citenamefont {Jones}\ and\ \citenamefont {Masanes}(2005)}]{JM.05}%
  \BibitemOpen
  \bibfield  {author} {\bibinfo {author} {\bibfnamefont {N.~S.}\ \bibnamefont
  {Jones}}\ and\ \bibinfo {author} {\bibfnamefont {L.}~\bibnamefont
  {Masanes}},\ }\href {\doibase 10.1103/PhysRevA.72.052312} {\bibfield
  {journal} {\bibinfo  {journal} {Phys. Rev. A}\ }\textbf {\bibinfo {volume}
  {72}},\ \bibinfo {pages} {052312} (\bibinfo {year} {2005})}\BibitemShut
  {NoStop}%
\bibitem [{\citenamefont {Bhatia}(1997)}]{Bha.97}%
  \BibitemOpen
  \bibfield  {author} {\bibinfo {author} {\bibfnamefont {R.}~\bibnamefont
  {Bhatia}},\ }\href@noop {} {\emph {\bibinfo {title} {Matrix Analysis}}},\
  Vol.\ \bibinfo {volume} {169}\ (\bibinfo  {publisher} {Springer},\ \bibinfo
  {year} {1997})\BibitemShut {NoStop}%
\bibitem [{\citenamefont {Navascu\'es}\ \emph {et~al.}(2007)\citenamefont
  {Navascu\'es}, \citenamefont {Pironio},\ and\ \citenamefont
  {Ac\'{\i}n}}]{NPA.07}%
  \BibitemOpen
  \bibfield  {author} {\bibinfo {author} {\bibfnamefont {M.}~\bibnamefont
  {Navascu\'es}}, \bibinfo {author} {\bibfnamefont {S.}~\bibnamefont
  {Pironio}}, \ and\ \bibinfo {author} {\bibfnamefont {A.}~\bibnamefont
  {Ac\'{\i}n}},\ }\href {\doibase 10.1103/PhysRevLett.98.010401} {\bibfield
  {journal} {\bibinfo  {journal} {Phys. Rev. Lett.}\ }\textbf {\bibinfo
  {volume} {98}},\ \bibinfo {pages} {010401} (\bibinfo {year}
  {2007})}\BibitemShut {NoStop}%
\bibitem [{\citenamefont {Navascu{\'{e}}s}\ \emph {et~al.}(2008)\citenamefont
  {Navascu{\'{e}}s}, \citenamefont {Pironio},\ and\ \citenamefont
  {Ac{\'{\i}}n}}]{NPA.08}%
  \BibitemOpen
  \bibfield  {author} {\bibinfo {author} {\bibfnamefont {M.}~\bibnamefont
  {Navascu{\'{e}}s}}, \bibinfo {author} {\bibfnamefont {S.}~\bibnamefont
  {Pironio}}, \ and\ \bibinfo {author} {\bibfnamefont {A.}~\bibnamefont
  {Ac{\'{\i}}n}},\ }\href {\doibase 10.1088/1367-2630/10/7/073013} {\bibfield
  {journal} {\bibinfo  {journal} {New Journal of Physics}\ }\textbf {\bibinfo
  {volume} {10}},\ \bibinfo {pages} {073013} (\bibinfo {year}
  {2008})}\BibitemShut {NoStop}%
\bibitem [{\citenamefont {Diamond}\ and\ \citenamefont {Boyd}(2016)}]{DB.16}%
  \BibitemOpen
  \bibfield  {author} {\bibinfo {author} {\bibfnamefont {S.}~\bibnamefont
  {Diamond}}\ and\ \bibinfo {author} {\bibfnamefont {S.}~\bibnamefont {Boyd}},\
  }\href@noop {} {\bibfield  {journal} {\bibinfo  {journal} {J. Mach. Learn.
  Res.}\ }\textbf {\bibinfo {volume} {17}},\ \bibinfo {pages} {2909–2913}
  (\bibinfo {year} {2016})}\BibitemShut {NoStop}%
\bibitem [{\citenamefont {Wehner}\ \emph {et~al.}(2008)\citenamefont {Wehner},
  \citenamefont {Doherty}, \citenamefont {Toner},\ and\ \citenamefont
  {Liang}}]{DW.08}%
  \BibitemOpen
  \bibfield  {author} {\bibinfo {author} {\bibfnamefont {S.}~\bibnamefont
  {Wehner}}, \bibinfo {author} {\bibfnamefont {A.~C.}\ \bibnamefont {Doherty}},
  \bibinfo {author} {\bibfnamefont {B.}~\bibnamefont {Toner}}, \ and\ \bibinfo
  {author} {\bibfnamefont {Y.}~\bibnamefont {Liang}},\ }in\ \href {\doibase
  10.1109/CCC.2008.26} {\emph {\bibinfo {booktitle} {Twenty-Third Annual IEEE
  Conference on Computational Complexity - CCC 2008}}}\ (\bibinfo  {publisher}
  {IEEE Computer Society},\ \bibinfo {address} {Los Alamitos, CA, USA},\
  \bibinfo {year} {2008})\ pp.\ \bibinfo {pages} {199--210}\BibitemShut
  {NoStop}%
\bibitem [{\citenamefont {Ac\'{\i}n}\ \emph {et~al.}(2012)\citenamefont
  {Ac\'{\i}n}, \citenamefont {Massar},\ and\ \citenamefont {Pironio}}]{AM.12}%
  \BibitemOpen
  \bibfield  {author} {\bibinfo {author} {\bibfnamefont {A.}~\bibnamefont
  {Ac\'{\i}n}}, \bibinfo {author} {\bibfnamefont {S.}~\bibnamefont {Massar}}, \
  and\ \bibinfo {author} {\bibfnamefont {S.}~\bibnamefont {Pironio}},\ }\href
  {\doibase 10.1103/PhysRevLett.108.100402} {\bibfield  {journal} {\bibinfo
  {journal} {Phys. Rev. Lett.}\ }\textbf {\bibinfo {volume} {108}},\ \bibinfo
  {pages} {100402} (\bibinfo {year} {2012})}\BibitemShut {NoStop}%
\bibitem [{\citenamefont {Moroder}\ \emph {et~al.}(2013)\citenamefont
  {Moroder}, \citenamefont {Bancal}, \citenamefont {Liang}, \citenamefont
  {Hofmann},\ and\ \citenamefont {G\"uhne}}]{MT.13}%
  \BibitemOpen
  \bibfield  {author} {\bibinfo {author} {\bibfnamefont {T.}~\bibnamefont
  {Moroder}}, \bibinfo {author} {\bibfnamefont {J.-D.}\ \bibnamefont {Bancal}},
  \bibinfo {author} {\bibfnamefont {Y.-C.}\ \bibnamefont {Liang}}, \bibinfo
  {author} {\bibfnamefont {M.}~\bibnamefont {Hofmann}}, \ and\ \bibinfo
  {author} {\bibfnamefont {O.}~\bibnamefont {G\"uhne}},\ }\href {\doibase
  10.1103/PhysRevLett.111.030501} {\bibfield  {journal} {\bibinfo  {journal}
  {Phys. Rev. Lett.}\ }\textbf {\bibinfo {volume} {111}},\ \bibinfo {pages}
  {030501} (\bibinfo {year} {2013})}\BibitemShut {NoStop}%
\end{thebibliography}%

\appendix

\section{Results for Section \ref{sec:3}}\label{ap:A}

\subsection{Extension of the region of interest} \label{ap:A1}

In the main text we have studied the quantum value of the functionals in Eq.~\eqref{eq:famber} assuming $\alpha_3\in[0, 2]$. Here we extend this analysis, for the first branch ($\alpha_2=0$), to all values of $\alpha_3$. Note first that, as stated in Ref.~\cite{Ka.20}, for the functional in $(\alpha_1=0, \alpha_3=2)$ the quantum value $\beta_Q$ coincides with the local value $\beta_L=4\alpha_3=8$, and it is not hard to see that this relation holds for $\alpha_3>2$. Indeed, let us denote by $W(\alpha_1, \alpha_3)$ the Bell operator associated to the member of the family in Eq.~\eqref{eq:famber} defined by the parameters $\alpha_1, \alpha_3$. Then for $\alpha_1=0$ and $\alpha_3=2+\epsilon$, with $\epsilon>0$, we can write
\begin{equation}
\begin{split}
    \mbraket{W(0, \alpha_3)}&=\;\mbraket{W(0, 2)}+\mbraket{W(0, \epsilon)} \\
    &\leq \;8+4\epsilon= \;4\alpha_3.
\end{split}
\end{equation}
Thus, for all the functionals along the $\alpha_1=0$ line, with $\alpha_3>2$, the local value is seen to be an upper bound for the quantum value, and therefore we conclude $\beta_Q=\beta_L=4\alpha_3$.

\begin{figure}[h]
\includegraphics[width=8.7cm,trim=0cm 0cm 0cm 0cm;clip]{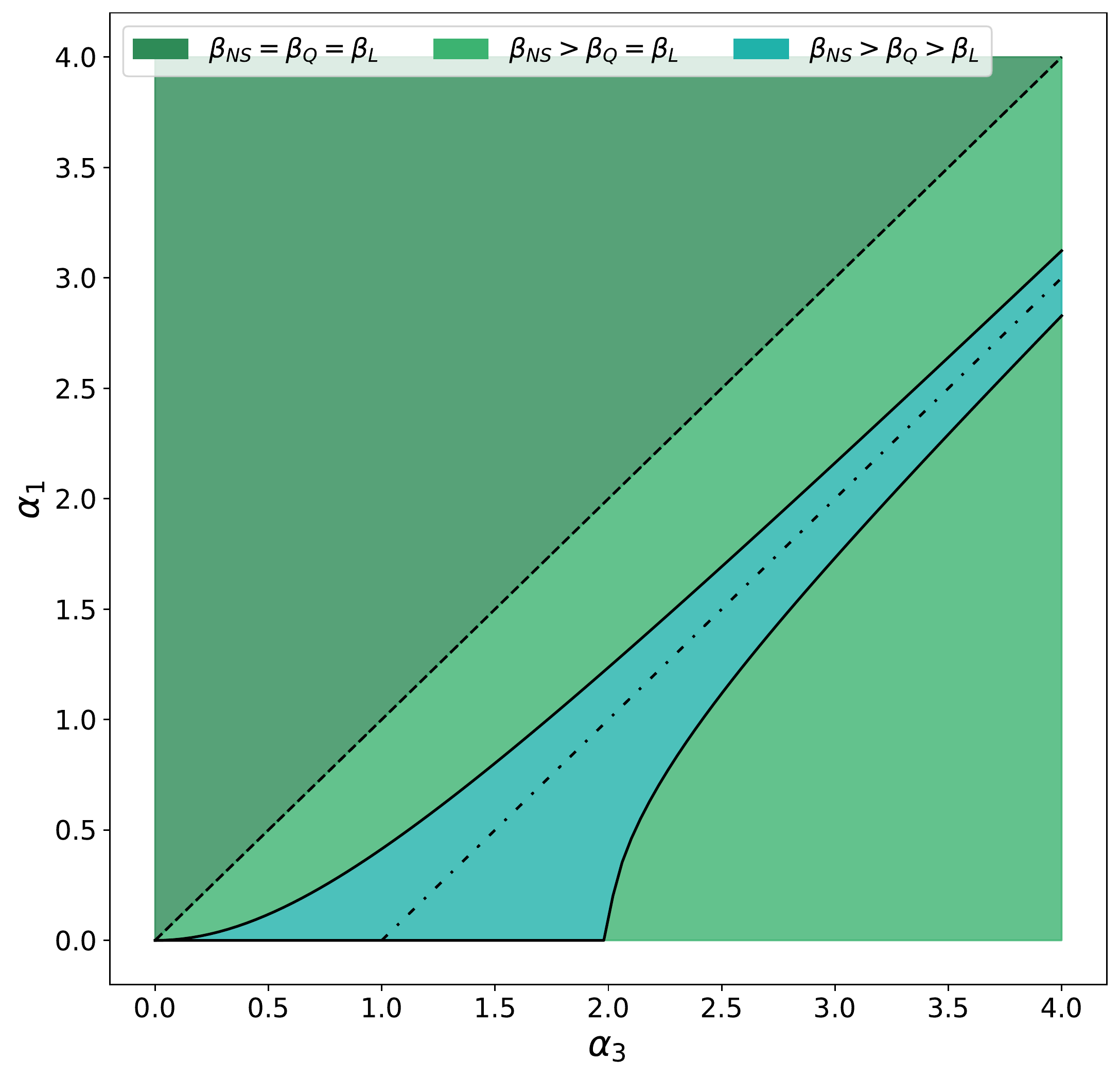}
\caption{Depiction of the quantum advantage ($\beta_Q>\beta_L$) region in the whole parameter space. The solid lines in the plot demarcate the boundary of this region, while the dashdotted line indicates the boundary between the region where the local value is $\beta_L=4\alpha_3$ ($\alpha_1<\alpha_3-1$) and that where $\beta_L=4(\alpha_1+1)$ ($\alpha_1>\alpha_3-1$). As in the previous plots, the region above the $\alpha_1=\alpha_3$ curve satisfies $\beta_{NS}=\beta_L$}\label{FIG:1A}
\end{figure}

Consider now the upper bound in Eq.~\eqref{eq:beta_Q} and the realisation attaining it, described in Eqs.~\eqref{eq:re1}-\eqref{eq:sinf}. The validity of this realisation is determined by Eq.~\eqref{eq:sint}, which implies
\begin{equation}
    0\leq\left(\gamma^2-\frac{\alpha_1^2}{\gamma^2} \right)\leq 2.
\end{equation}
A direct calculation shows that this constraint implies, if we consider $\alpha_3>2$,
\begin{equation}
    \sqrt{\alpha_3(\alpha_3-2)}\leq\alpha_1\leq\sqrt{\alpha_3^2+1}-1.\label{eq:bdary}
\end{equation}
These inequalities define the region in parameter space where the given realisation is valid, which is shown in FIG.~\ref{FIG:1A}. This region is clearly unbounded since the curves defining its boundaries do not intersect. Actually, these curves are respectively below and above the $\alpha_1=\alpha_3-1$ line which is in turn the boundary between the regions depicted in FIG.~\ref{FIG:1} where the local value is either $\beta_L=4\alpha_3$ or $\beta_L=4(\alpha_1+1)$. The results derived in the main text are easily seen to extend to all values of $\alpha_3$ for all functionals above the curve $\alpha_1=\sqrt{\alpha_3(\alpha_3-2)}$. On the curve, moreover, the quantum value coincides with the local value $\beta_L=4\alpha_3$, as can be checked directly using Eqs.~\eqref{eq:re1}-\eqref{eq:sinf}. From here we can conclude that below this curve the relation $\beta_Q=\beta_L=4\alpha_3$ also holds. Indeed, note that given $\alpha_1 <\sqrt{\alpha_3(\alpha_3-2)}$ and $\alpha_3>2$ we can write the ensuing Bell operator as
\begin{equation}
    W(\alpha_1, \alpha_3)= p\, W(\tilde\alpha_1, \alpha_3)+(1-p)\, W(0, \alpha_3)\leq 4\alpha_3
\end{equation}
for some $p\in(0, 1)$, where $\tilde\alpha_1=\sqrt{\alpha_3(\alpha_3-2)}$, and where in the last inequality we used that $W(\tilde\alpha_1, \alpha_3)\leq4\alpha_3$ and $W(0, \alpha_3)\leq 4\alpha_3$ as we have shown above. With these results we complete the study of the quantum value and the realisations attaining it for all functionals in the branch $\alpha_2=0$.

\subsection{Projectivity of measurements} \label{ap:A2}

In the following we show that, for the functionals within the region of quantum advantage depicted in FIG \ref{FIG:3}, the quantum value $\beta_Q$ can only be attained if observables $A_{1(2)}$ and $B_{1(2)}$ describe projective measurements.

If we square the polynomials $P_i$ in the SOS decomposition \eqref{eq:sos} without assuming $A_x^2=B_y^2=\mathbb{1}$ we get
\begin{equation}\label{eq:sumi}
\begin{split}
    \sum_{i=1}^3 P_i^2=& \quad 2\gamma^2 \mathbb{1}\otimes\mathbb{1} - W \\
    &+ \frac{\alpha_3^2}{4\gamma^2}\left[ (A_1+A_2)^2\otimes\mathbb{1}+\mathbb{1}\otimes(B_1+B_2)^2\right]  \\
    &+ \frac{\alpha_3^2}{4\gamma^2}\left[ (A_1-A_2)^2\otimes B_3^2+A_3^2\otimes(B_1-B_2)^2\right].
\end{split}
\end{equation}
To upper bound the terms in the last line we use the fact that for all measurements the following operator inequalities hold: $(A_1-A_2)^2\otimes B_3^2\leq (A_1-A_2)^2\otimes \mathbb{1}$ and $A_3^2\otimes(B_1-B_2)^2\leq \mathbb{1}\otimes(B_1-B_2)^2$. As a consequence we obtain
\begin{eqnarray}
    \sum_{i=1}^3 P_i^2\leq&& 2\gamma^2 \mathbb{1}\otimes\mathbb{1} - W \label{eq:ineq1}\\
    &+&\frac{\alpha_3^2}{2\gamma^2}\left[(A_1^2+A_2^2)\otimes\mathbb{1}+\mathbb{1}\otimes(B_1^2+B_2^2)\right] \notag \\
    \leq&&\beta_Q \mathbb{1}\otimes\mathbb{1} - W, \label{eq:ineq2}
\end{eqnarray}
where in the last line we used $A_x^2\leq\mathbb{1}$ and $B_y^2\leq\mathbb{1}$.

Consider now a realisation $\{A_x, B_y, \rho_{AB}\}$ such that $\Tr\, W\rho_{AB}=\beta_Q$, which in virtue of these inequalities implies $\Tr\,P_i^2=0\,\forall i$.  Then from inequality \eqref{eq:ineq2} follows that $\Tr\, A_{1(2)}^2\rho_A=\Tr\, B_{1(2)}^2\rho_B=1$, which implies $A_{1(2)}^2=B_{1(2)}^2=\mathbb{1}$ if, as in the main text, marginals of $\rho_{A(B)}$ are assumed to be full-rank. Projectivity of $A_3$ and $B_3$ is for the moment still assumed, and we will prove it in next section.

\subsubsection{Derivation of \texorpdfstring{Eqs.~\eqref{eq:str1} and \eqref{eq:str2}}{}} \label{ap:A2a}

Here we derive relations \eqref{eq:str1} and \eqref{eq:str2} from the main text. The realisation is assumed to attain the quantum value $\beta_Q$, and this implies the support of the state $\rho_{AB}$ is included in the intersection of the kernels of the polynomials $\{P_i\}$ giving the SOS decomposition \eqref{eq:sos}, i.e., $P_i\rho_{AB}=0\,\forall\, i$. From $P_1\rho_{AB}=0$ we get
\begin{equation}
    [\alpha_1 (A_1+A_2)\otimes \mathbb{1}+ \alpha_3(A_1-A_2)\otimes B_3]^2\rho_{AB}=4\gamma^4\rho_{AB}. \label{eq:e1s}
\end{equation}
Expanding the square on the left hand side gives
\begin{equation}
    [2(\alpha_3^2+\alpha_1^2)\mathbb{1}-2\gamma^2\{A_1, A_2\}]\otimes\mathbb{1}\rho_{AB}=4\gamma^4\rho_{AB},
\end{equation}
which leads, after tracing out Bob's system, to
\begin{equation}
    \{A_1, A_2\}\rho_A=2\left[ 1-\gamma^2+\frac{\alpha_1^2}{\gamma^2}  \right]\rho_A. 
\end{equation}
Since $\rho_A$ is assumed to have full-rank we can right-multiply by $\rho_A^{-1}$ to get
\begin{equation}
    \{A_1, A_2\}=2\left[ 1-\gamma^2+\frac{\alpha_1^2}{\gamma^2}  \right]\mathbb{1}. \label{eq:str1A}
\end{equation}
Starting from $P_2\rho_{AB}=0$ and following an analogous argument we can arrive at the same equation for the anticommutator $\{B_1, B_2\}$.

Now consider again $P_1\rho_{AB}=0$, which implies
\begin{equation}
    [\alpha_1 (A_1+A_2)\otimes \mathbb{1}+\alpha_3 (A_1-A_2)\otimes B_3]\rho_{AB}=2\gamma^2\rho_{AB}, \label{eq:e2s}
\end{equation}
and use $P_3\rho_{AB}=0$ to write
\begin{equation}
(A_1+A_2)\otimes\mathbb{1}\rho_{AB}=\mathbb{1}\otimes(B_1+B_2)\rho_{AB},   
\end{equation}
which we can plug in Eq.~\eqref{eq:e2s} to get
\begin{equation}
    [\alpha_1\mathbb{1}\otimes(B_1+B_2)+\alpha_3(A_1-A_2)\otimes B_3]\rho_{AB}=2\gamma^2\rho_{AB}.
\end{equation}
Squaring this relation and using the result in Eq.~\eqref{eq:str1A} we arrive at
\begin{equation}
    (A_1-A_2)\otimes\{B_1+B_2, B_3\}\rho_{AB}=0.
\end{equation}
It is easy to see that Eq.~\eqref{eq:str1A} also implies $(A_1-A_2)^2\propto\mathbb{1}$. Then, multiplying by $(A_1-A_2)\otimes\mathbb{1}$ we get
\begin{equation}
    \mathbb{1}\otimes\{B_1+B_2, B_3\}\rho_{AB}=0.
\end{equation}
Finally, if we trace out Alice's system and use that $\rho_B$ is full-rank we arrive at
\begin{equation}
\{B_1+B_2, B_3\}=0 \label{eq:str2A}.
\end{equation}
As before, an analogous procedure shows that relation \eqref{eq:str2A} is also satisfied by Alice's observables.

We can now consider operator $B_3$. Note that in an optimal realisation this operator is completely determined by maximising the average $\Tr (A_1-A_2)\otimes B_3\rho_{AB}=\Tr X A_3$, with 
\begin{equation}
    X=\sqrt{\rho_A}(A_1-A_2)\sqrt{\rho_A}.
\end{equation}
Writing now $B_3=M^B_{1|3} - M^B_{-1|3}$ it is seen that $M^B_{1(-1)|3}$ is a projector onto the positive (negative) eigenspace of $X$, which implies $B_3^2=\mathbb{1}$ if $X$ is full-rank. This last condition is ensured by Eq.~\eqref{eq:str1A}, since it implies $(A_1-A_2)^2\propto\mathbb{1}$, and then that $(A_1-A_2)$ is full-rank. The projectivity of $B_3$ is proven in an analogous way.

\subsection{General forms of local observables} \label{ap:A3}

The anticommutation relations we derived in the previous section can now be used to find the general form of the local measurement operators in the optimal realisation. From Jordan's lemma we know that there is a basis of the local Hilbert space $\HC_A$ in which observables $A_1$ and $A_2$ take a jointly block-diagonal form, with block dimension being either $1$ or $2$. Condition $A_{1(2)}^2=\mathbb{1}$ implies that the coefficients in one-dimensional blocks must be $\pm 1$, which is incompatible with relation \eqref{eq:str1A} inside the quantum advantage ($\beta_Q>\beta_L$) region. This relation thus implies that all blocks are two dimensional and the principal angles between $A_1$ and $A_2$ are all the same. Therefore we can write $\HC_A=\mathbb{C}^2\otimes\mathbb{C}^d$ for some $d\in\mathbb{N}$ and, up to a unitary,
\begin{equation}
\begin{split}
    A_1&=[\cos(\theta)\sigma_z+\sin(\theta)\sigma_x]\otimes\mathbb{1}, \\
    A_2&=[\cos(\theta)\sigma_z-\sin(\theta)\sigma_x]\otimes\mathbb{1},
\end{split}
\end{equation}
where the angle $\theta$ is related to $\alpha_1$ and $\alpha_3$ by relation \eqref{eq:sint}.

We can now move on to finding $A_3$. From relation \eqref{eq:str2A} (on Alice's side) we get that $A_3$ should exhibit the same block structure as $A_1+A_2$, and then the anticommutator should vanish blockwise. Thinking of each block in the structure as an independent set of two-qubit observables, we can translate the vanishing anticommutator into an orthogonality condition between the direction defined by $A_3$ in the Bloch sphere and that defined by $A_1+A_2$. We then see (this problem is fully worked out in Ref.~\cite{Ka.20}) that the most general form for $A_3$ is
\begin{equation}
    A_3=\sum_k[\cos(\mu_k)\sigma_x+\sin(\mu_k)\sigma_y]\otimes\ketbra{k}{k}
\end{equation}
where $\mu_k\in[0, 2\pi]$ and $\{\ket{k}\}$ is an orthonormal basis in $\mathbb{C}^d$. Note that this angle is not required to be necessarily the same across the different blocks, as it is the case with $\theta$ in the expressions for $A_1$ and $A_2$.

The same relations we used to characterise the observables in Alice's side hold for Bob's observables as well, so we might just assume the same expressions for $B_1$, $B_2$ and $B_3$ respectively. It is worth noting, however, that in the functional \eqref{eq:famber} $B_3$ only couples to $A_1-A_2$ and $A_3$ to $B_1-B_2$. It is therefore convenient, to rotate the basis in Bob's side and write his observables as
\begin{equation*}
\begin{split}
    B_1&=\sum_k[\cos(\theta)\sigma_z+\sin(\theta)(\cos(\nu_k)\sigma_x-\sin(\nu_k)\sigma_y)]\otimes\ketbra{k}{k},\\
    B_2&=\sum_k[\cos(\theta)\sigma_z-\sin(\theta)(\cos(\nu_k)\sigma_x-\sin(\nu_k)\sigma_y)]\otimes\ketbra{k}{k}, \\
    B_3&=\sigma_x\otimes\mathbb{1},
\end{split}
\end{equation*}
where $\nu_k\in[0, 2\pi]$.

\subsubsection{Optimal two-qubit realisation} \label{ap:A3a}

We have found the general form of the local observables in the optimal realisation, and now we can use it to build the associated Bell operator, which takes the form
\begin{equation}
    W=\sum_{k, k'}R(\mu_k, \nu_{k'})\otimes\ketbra{k}{k}\otimes\ketbra{k'}{k'}, %\label{eq:RA}
\end{equation}
where $R(\mu_k, \nu_{k'})$ is the $2$-qubit Bell operator built up with the non-trivial part of the operators $\{A_x, B_y\}$ given above, and coincides with the Bell operator when the local dimension is $2$. We will now focus on this case.

Before going any further it is convenient to go back to the polynomials defining the SOS decomposition \eqref{eq:sos}, and note that for a realisation attaining the optimal value $\beta_Q$ the equation $P_3\rho_{AB}=0$ implies
\begin{equation}
    (A_1+A_2)\otimes(B_1+B_2)\rho_{AB}=\mathbb{1}\otimes(B_1+B_2)^2\rho_{AB}. \label{eq:eigr}
\end{equation}
Relation \eqref{eq:eigr} holds for any state $\rho_{AB}$ of a realisation attaining the quantum value $\beta_Q$, in particular for a pure state $\ket{\Phi}$ such that $W\ket{\Phi}=\beta_Q\ket{\Phi}$. In such case $\rho_{AB}$ is the projector onto the pure state $\ket{\Phi}$ and it follows from Eq.~\eqref{eq:eigr} that the latter must be an eigenstate with $+1$ eigenvalue of $(A_1+A_2)\otimes(B_1+B_2)$, which for a two dimensional realisation implies
\begin{equation}
    \ket{\Phi}=\cos\left(\frac{\varphi}{2}\right)\ket{00}+e^{i\eta}\sin\left(\frac{\varphi}{2}\right)\ket{11} \label{eq:eigpsi}
\end{equation}
for some parameters $\varphi$ and $\eta$. We can now compute the average of $R(\mu_k, \nu_{k'})$ with the state \eqref{eq:eigpsi} and find
\begin{eqnarray}
    \mbraket{R(\mu_k, \nu_{k'})} &=& \alpha_1\cos(\theta)\cos(\varphi)+4\cos^2(\theta)+  \label{eq:aveg}\\
    &&2\alpha_3\sin(\theta)\sin(\varphi)[\cos(\eta)+\cos(\eta+\nu_{k'}-\mu_k)]. \notag
\end{eqnarray}
The quantum value $\beta_Q$ is the maximal value attainable in Eq.~\eqref{eq:aveg}, which implies $\eta=0$ in Eq.~\eqref{eq:eigpsi} and $\mu_k=\nu_{k'}$ in $R(\mu_k, \nu_{k'})$. This reduces the average \eqref{eq:aveg} to
\begin{equation}
    \mbraket{W}=4[\alpha_1\cos(\theta)\cos(\varphi)+\alpha_3\sin(\theta)\sin(\varphi)]+4\cos^2(\theta),
\end{equation}
and it is easy to check that this expression takes the value $\beta_Q$ iff relations \eqref{eq:cosf} and \eqref{eq:sinf} are fulfilled, implying the solution is unique. 

The results derived above show that, up to local unitaries, a two-qubit realisation can attain the optimal value $\beta_Q$ only with the state
\begin{equation}
    \ket{\Psi}=\cos\left(\frac{\varphi}{2}\right)\ket{00}+\sin\left(\frac{\varphi}{2}\right)\ket{11}
\end{equation}
and with observables
\begin{equation*}
\begin{split}
    A_1&=\cos(\theta)\sigma_z+\sin(\theta)\sigma_x, \\
    A_2&=\cos(\theta)\sigma_z-\sin(\theta)\sigma_x,\\
    A_3&=\cos(\mu)\sigma_x+\sin(\mu)\sigma_y, \\
    B_1&=\cos(\theta)\sigma_z+\sin(\theta)(\cos(\mu)\sigma_x-\sin(\mu)\sigma_y),\\
    B_2&=\cos(\theta)\sigma_z-\sin(\theta)(\cos(\mu)\sigma_x-\sin(\mu)\sigma_y), \\
    B_3&=\sigma_x,
\end{split}
\end{equation*}
with $\mu\in[0, 2\pi]$.

\section{Equivalent form of \texorpdfstring{$I_{3322}$}{} functional and \texorpdfstring{$\PV^{(n)}$}{} realisations}\label{ap:B}

The family of realisations we denote by $\PV^{(n)}$ in the main text was first introduced in Ref.~\cite{PV.10} in a slightly different manner. The functional we are optimising in Section \ref{sec:PV} is, for $\alpha_1=1$, $\alpha_3=1$:
\begin{equation}
\begin{split}
    \beta := &\;\quad \mbraket{A_1}+\mbraket{A_2} +\mbraket{B_1} +\mbraket{B_2} \\
    & -\mbraket{A_1 B_1} -\mbraket{A_1 B_2} -\mbraket{A_2 B_1} -\mbraket{A_2 B_2} \\
    &+ \mbraket{A_3 B_1} - \mbraket{A_3 B_2} + \mbraket{A_1 B_3} - \mbraket{A_2 B_3}.
\end{split}
\end{equation}
Since $A_1\leftrightarrow A_2, B_3\rightarrow -B_3$ and $B_1\leftrightarrow B_2, A_3\rightarrow -A_3$ are symmetries of the functional we can write, after flipping the sign of $B_1+B_2$,
\begin{equation}
\begin{split}
    \beta:= &\;\quad \mbraket{A_1}+\mbraket{A_2} -(\mbraket{B_1}+\mbraket{B_2}) \\
    & +\mbraket{A_1 B_1} +\mbraket{A_1 B_2} +\mbraket{A_2 B_1} +\mbraket{A_2 B_2} \\
    &+\mbraket{A_3 B_2} - \mbraket{A_3 B_1} + \mbraket{A_2 B_3} - \mbraket{A_1 B_3}.
\end{split}
\end{equation}
Furthermore, after writing $A_x=M^A_{1|x}-M^A_{-1|x}=2M^A_{1|x}-\mathbb{1}$ and $B_y=M^B_{1|y}-M^B_{-1|y}=2M^B_{1|y}-\mathbb{1}$ this functional becomes
\begin{equation}
\begin{split}
    I_{3322}=4&-\mbraket{M^A_{1|2}}-\mbraket{M^B_{1|1}}-2\mbraket{M^B_{1|2}} \label{eq:3322pv}\\
    &+\mbraket{M^A_{1|1}M^B_{1|1}}+\mbraket{M^A_{1|1}M^B_{1|2}} \\
    &+\mbraket{M^A_{1|2}M^B_{1|1}}+\mbraket{M^A_{1|2}M^B_{1|2}}  \\
    &-\mbraket{M^A_{1|1}M^B_{1|3}}+\mbraket{M^A_{1|2}M^B_{1|3}} \\
    &-\mbraket{M^A_{1|3}M^B_{1|1}}+\mbraket{M^A_{1|3}M^B_{1|2}}.
\end{split}
\end{equation}
Ignoring the constant term, Eq.~\eqref{eq:3322pv} is the form of the $I_{3322}$ functional studied in Ref.~\cite{PV.10}. Note that the authors define the observables as having spectrum $\{0, 1\}$. This choice leads to $A_x, (B_y)$ being the projector onto one of the outcomes, which in our notation would be $M^{A(B)}_{1|x(y)}$. As a result of these transformations, it is clear that the observables $A_{1, 2}^{(n)}$ we introduced in Eqs. \eqref{eq:A12} and \eqref{eq:A3} are linked to those presented in Ref.~\cite{PV.10} as $\hat A_{1, 2}$ by the relation $A_x=2\hat A_x-\mathbb{1}$, $x=1, 2$. In the case of observables $B_{1, 2}^{(n)}$ the same relation holds with respect to $\hat B_{1, 2}$, but in addition we flip the sign of $B_1+B_2$, which can be done by taking $\theta^B_j\rightarrow -\theta^B_j+\pi$. On the other hand, it is clear that $A_3^{(n)}$ has the same form as $\hat A_3$ in Ref.~\cite{PV.10} after the basis order is reversed, being the same true for $B_3^{(n)}$ and $\hat B_3$. Finally, note that the state $\ket{\Psi^{(n)}}$ can be obtained from that given in Ref.~\cite{PV.10} by reversing the order of the Schmidt basis on Bob's side.

\end{document}